\documentclass[preprint]{elsarticle}
\usepackage{ntheorem}
\newtheorem{hyp}{Hypothesis}

\makeatletter
\newcounter{subhyp}
\let\savedc@hyp\c@hyp

\newcommand{\normhyp}{%
  \let\c@hyp\savedc@hyp
  \renewcommand\thehyp{\arabic{hyp}}%
}
\makeatother

\usepackage{multicol}
\usepackage{color}
\usepackage{xspace}
\usepackage{pdfwidgets}
\usepackage{enumerate}
\usepackage{natbib}
\usepackage{amsmath,amssymb,amsfonts}
\usepackage{algorithmic}
\usepackage{graphicx}
\usepackage{subcaption}
\usepackage{tabularx}
\usepackage{textcomp}
\usepackage{lmodern}
\usepackage[table]{xcolor}
\usepackage[detect-all, separate-uncertainty]{siunitx}
\usepackage{todonotes}
\usepackage{enumitem}
\usepackage{bbding}
\usepackage{pifont}
\usepackage{multirow}
\usepackage{orcidlink}
\usepackage{listings}
\usepackage{comment}
\usepackage{xspace}
\usepackage{booktabs}
\usepackage[autostyle=false, style=english]{csquotes}
\usepackage[normalem]{ulem}
\usepackage{pgfplots}
\usepackage{textcomp}
\usepackage{tikz}
\usetikzlibrary{positioning,arrows.meta}
\definecolor{lightgray}{gray}{0.9}

\newcolumntype{Y}{>{\raggedright\arraybackslash}X}
 
\newcommand{\GOK}{{\textsc{InstruBPM}}\xspace}
\newcommand{\OLDGOK}{{\textsc{BPMG-IT}}\xspace}

\makeatletter
\def\bs{\expandafter\@gobble\string\\}
\def\lb{\expandafter\@gobble\string\{}
\def\rb{\expandafter\@gobble\string\}}
\def\@pdfauthor{Gökberk Çelikmasat et al.}
\def\@pdftitle{Generating BPMN Models with LLMs using Instruction Tuning}
\def\@pdfsubject{Document formatting with elsarticle.cls}
\def\@pdfkeywords{LaTeX, Elsevier Ltd, document class}

\DeclareRobustCommand{\LaTeX}{L\kern-.26em%
        {\sbox\z@ T%
         \vbox to\ht\z@{\hbox{\check@mathfonts
           \fontsize\sf@size\z@
           \math@fontsfalse\selectfont
          A\,}%
         \vss}%
        }%
     \kern-.15em%
    \TeX}
\makeatother

\pgfplotsset{compat=1.18}

\begin{document}

\def\testa{This is a specimen document. }
\def\testc{\testa\testa\testa\testa}
\def\testb{\testc\testc\testc\testc\testc}
\long\def\test{\testb\par\testb\par\testb\par}

\pinclude{\copy\contbox\printSq{\LastPage}}

\begin{frontmatter}

\title{\GOK: Instruction-Tuning Open-Weight Language Models for BPMN Model Generation}

\author[1]{Gökberk Çelikmasat\corref{cor1}}
\ead{gokberk.celikmasat@std.bogazici.edu.tr}
\affiliation[1]{organization={Boğaziçi University},
                city={İstanbul},
                country={Türkiye}}

\author[1]{Atay Özgövde}
\ead{ozgovde@bogazici.edu.tr}

\author[2]{Fatma Başak Aydemir}
\ead{f.b.aydemir@uu.nl}
\affiliation[2]{organization={Utrecht University},
                city={Utrecht},
                country={The Netherlands}}

\cortext[cor1]{Corresponding author}

\date{\today}

\begin{abstract}
Business process models capture how work flows through an organization and are central to business process management. 
Yet, despite their value, practitioners often skip modeling because it is time-consuming and demands scarce expertise.
We address this barrier by investigating whether open-weight large language models, adapted via instruction tuning, can generate high-quality BPMN process models directly from natural language descriptions in a cost-effective and privacy-preserving way. We introduce \GOK, a reproducible approach that prepares paired text–diagram data and instruction tunes an open source large language model with parameter-efficient fine-tuning and quantization for on-prem deployment. We evaluate the tuned model through complementary perspectives: (i)  text/code similarity using BLEU, ROUGE-L, and METEOR, (ii) structural fidelity using Relative Graph Edit Distance, (iii)  guidelines conformance using external tool checks, and (iv) a small expert review.
Using a curated subset of a multi-domain BPMN dataset, we compare the tuned model with untuned open-weight baselines and strong proprietary models under consistent prompting regimes. Our compact, tuned model outperforms untuned open-weight and proprietary baselines across sequence and structural metrics, and matches or surpasses its substantially larger predecessor while requiring far fewer resources; guideline analysis and expert feedback further indicate that the generated diagrams largely follow BPMN best practices and serve as useful starting points that reduce modeling effort.
Overall, instruction tuning improves structural accuracy and robustness compared to untuned baselines and reduces reliance on heavy prompt scaffolding. We publicly share the trained models and scripts to support reproducibility and further research.
\end{abstract}


\end{frontmatter}

\section{Introduction}
\label{sec:intro}

Business process models (BPMs) describe the sequence of tasks, events, and decisions that define how work is conducted within organizations, playing a central role in business process management and digital transformation initiatives~\cite{pihir2019business}. Among modeling languages, the Business Process Model and Notation (BPMN) has become the de facto standard due to its rich graphical vocabulary and wide adoption across both industry and academia~\cite{geiger2018bpmn}. BPMN 2.0~\cite{2013international} provides a standardized graphical language for describing processes in a way that is both machine-readable and understandable to business stakeholders.
By providing standardized graphical constructs for activities, events, gateways, and sequence flows, BPMN supports clear and consistent process documentation.

Despite this widespread adoption, creating high-quality BPMN models manually remains a complex and demanding task. An empirical analysis of 585 BPMN 2.0 process models from six companies by Leopold et al.~\cite{leopold2016learning} reports recurring quality issues in split/join representations, message-flow usage, model decomposition, and consistent labeling, reflecting both the notation's complexity and the difficulty practitioners face in mastering its representational choices.

Existing automation approaches, including rule-based methods, machine learning (ML) pipelines, and natural language processing (NLP) techniques, offer partial support but often lack scalability, structural fidelity, or the ability to generalize across diverse process descriptions~\cite{sonbol2023machine, honkisz2018concept, van2018challenges}.

Recent generative-AI work relevant to automated model generation has explored zero-shot prompting, CoT, ToT, thinking-style decoding, meta-prompting, automatic prompt optimization, and retrieval-augmented generation~\cite{silva2024application,li2024llm,wu2024thinking,suzgun2024meta,li2025survey,minor2024retrieval}. Within BPMN and process-modeling research specifically, prompt-based approaches are also prominent~\cite{kourani2024process}. These methods adapt behavior through the context window rather than model parameters, which can increase prompt complexity, inference cost, and latency. Weight-level adaptation has therefore emerged as a promising alternative in text-to-BPMN work~\cite{licardo2024method,nivon2024automated,nivon2024semi}. Building on recent fine-tuning-based BPMN generation studies~\cite{soliman2025size,li2025llmbpmn}, we study whether a compact open-weight model can be instruction-tuned for BPMN generation under realistic on-premise constraints.

To address these challenges, we propose the \GOK approach to study whether instruction-tuned, open-weight LLMs can generate structurally correct BPMN models directly from natural language descriptions. Using the MaD dataset~\cite{li2023mad} of paired natural language descriptions and BPMN models captured in DOT language, we instruction-tune Qwen3-4B~\cite{yang2025qwen3} with parameter-efficient fine-tuning (PEFT)~\cite{han2024parameter} and evaluate outputs along four aspects:
\begin{itemize}
  \item textual similarity (BLEU, ROUGE-L, METEOR);
  \item structural fidelity via Relative Graph Edit Distance (R-GED);
  \item practical quality using BEBoP (understandaBility vErifier for 
  Business Process models)~\cite{corradini2018guidelines} guideline checks; and
  \item a small expert review.
\end{itemize}
Our approach thus uses instruction tuning to embed domain constraints directly into the model parameters, offering a more efficient, self-contained solution for on-premises deployment that does not depend on external scaffolding at inference time.

This article extends our prior work \emph{\OLDGOK}~\cite{ccelikmasat2025generating_profes}, in which we instruction-tuned a single open-weight 9B-parameter model via QLoRA~\cite{dettmers2023qlora} and evaluated it on a diverse but relatively small set of 45 textual process descriptions using BLEU, ROUGE-L, METEOR, Graph Edit Distance metrics, and a qualitative expert study. While those results provided initial evidence that instruction tuning improves both textual similarity and structural correctness over off-the-shelf LLMs, the study did not yet consider guideline conformance, deployment-oriented trade-offs, or ablation analyses. Compared to this preliminary study, our present research makes three main contributions:  

\begin{itemize}
  \item {A redesigned artifact around a compact Qwen3-4B backbone and the \emph{\GOK} method}, yielding a reproducible instruction-tuning pipeline and a family of quantized variants tailored to on-prem deployment.
  
  \item {A broadened evaluation on a stratified set of 180 descriptions across 15 business domains} - exactly four times the 45-instance test set used in our prior work - together with a re-evaluation of the Gemma2-9B-\OLDGOK baseline under this new protocol. We also add new evaluation lenses, including R-GED, BEBoP guideline checks, statistical tests, and an updated expert assessment, resulting in a multi-perspective evaluation that couples text/graph metrics with guideline and practitioner judgments.
  
  \item {A systematic investigation of deployment-relevant design choices}, including LoRA configuration, merge-time $\alpha$-scaling, and quantization, through controlled ablation studies that yield empirical insights about model and infrastructure trade-offs directly relevant for practitioners in on-premises settings.
\end{itemize}

Compact base models are particularly attractive in this setting, where organizations increasingly need to run private modeling assistants alongside other workloads on a single accelerator without compromising on quality. On the 180-instance stratified benchmark introduced below, our resulting compact pipeline matches or slightly surpasses Gemma2-9B-\OLDGOK on all four headline metrics (BLEU, ROUGE-L, METEOR, and R-GED accuracy; see Table~\ref{table:bleu-rouge}) while only using less than half (${\sim}45\%$) of the active parameters and memory.

The remainder of this paper is organized as follows. Section \ref{sec:background} provides an overview of core concepts used in this paper, and Section \ref{sec:rel_work} reviews related work on BPMN generation and LLM adaptation. Section \ref{sec:res-method} introduces the research method and hypotheses. Section \ref{sec:BPMN_generation_methodology} presents the \GOK method. Section \ref{sec:results_and_analysis} reports evaluation results with respect to our hypotheses. Section \ref{sec:discussion} discusses findings, implications, and limitations. Finally, Section \ref{sec:conclusion_and_future_work} concludes with future directions.

\section{Background}
\label{sec:background}

This section introduces foundational concepts on BPMN and automated model generation, instruction tuning, prompt engineering, and inference-time reasoning strategies.

\subsection{BPMN and Automated Model Generation}
\label{bpmn_background}

BPMN 2.0~\cite{2013international} is an internationally standardized graphical language (ISO/IEC 19510) for representing business processes. BPMN facilitates communication among stakeholders by visually capturing workflows, tasks, events, and gateways, while maintaining machine-readability for automated analysis. These properties make BPMN the preferred notation for process modeling in both industry and research.

Automating the generation of BPMN models from textual descriptions has long been a research challenge~\cite{friedrich2011process}. The task involves interpreting unstructured natural language, identifying process elements, and translating them into a structured, diagrammatic form. Traditional approaches have explored both rule-based extraction pipelines and machine learning techniques. While these methods demonstrate partial success, they often struggle with scalability, ambiguity resolution, and structural consistency, motivating the exploration of more powerful generative techniques.

\subsection{Instruction Tuning}
\label{instr_tune_background}
Instruction tuning is a form of supervised fine-tuning in which each training example pairs an explicit natural-language instruction with a target output, rather than only an input-output example as in conventional fine-tuning~\cite{wei2021finetuned}. Conventional fine-tuning adjusts model weights toward a single task format implied by the training distribution, whereas instruction tuning conditions the model on the instruction itself, enabling the same model to learn to follow a wide range of instructions and generalize to unseen ones. In contrast to prompt engineering, which adapts behavior through the context window without changing parameters, instruction tuning internalizes the task into the model weights through supervised optimization.

Concretely, a training example might pair an instruction such as ``Draft a formal email requesting an extension for a project deadline'' with the corresponding email as the target output; supervised loss is then computed against that output, conditioned on the instruction. The training process employs these instruction-output pairs to minimize supervised loss functions, guiding the model's internal parameters toward accurate response generation.

Sanh \emph{et al.}~\cite{sanh2022multitask} and Wei \emph{et al.}~\cite{wei2021finetuned} show that instruction tuning substantially improves zero-shot task generalization; in the FLAN study, the instruction-tuned model outperformed zero-shot GPT-3 on 20 of the 25 evaluated tasks~\cite{wei2021finetuned}. These findings establish instruction tuning as an effective mechanism for improving instruction following on unseen tasks and motivate its use here for structured BPMN generation.

However, full fine-tuning of LLMs with billions of parameters is often computationally expensive and can be impractical under limited GPU memory or when multiple domain variants must be trained. PEFT~\cite{han2024parameter} reduces this cost by updating only a small subset of trainable parameters while keeping the base model weights frozen. A widely used PEFT technique is \emph{LoRA} (Low-Rank Adaptation)~\cite{hu2022lora}, which represents the update to a target weight matrix $W \in \mathbb{R}^{d_{\text{out}} \times d_{\text{in}}}$ as a low-rank factorization:
\begin{equation}
\Delta W \;=\; \frac{\alpha}{r}\, A B^\top,
\label{eq:lora}
\end{equation}
where $d_{\text{in}}$ and $d_{\text{out}}$ denote the input and output dimensions of the adapted weight matrix, respectively, and $r$ denotes the LoRA rank. The low-rank factors are $A \in \mathbb{R}^{d_{\text{out}} \times r}$ and $B \in \mathbb{R}^{d_{\text{in}} \times r}$. The effective weight at inference or after merging the adapter into the base model is $W^\star = W + \Delta W$. The rank $r$ determines the number of trainable parameters introduced by LoRA and thus the adaptation capacity, while the scaling factor $\alpha$ controls the magnitude of the injected update. In practice, merge-time scaling can be used to adjust the degree to which the learned adapter influences the merged checkpoint: very small $\alpha$ values may under-apply the adaptation, while very large values may over-apply it and degrade generalization.

Zhang \emph{et al.}~\cite{zhang2023instruction} emphasize that combining PEFT with instruction tuning allows LLMs to rapidly adapt to specialized domains with minimal retraining. This is especially valuable in applications like BPMN modeling, where accurately interpreting domain-specific instructions is essential. 

Beyond training-time efficiency, the memory footprint can be further reduced at deployment by using quantization, which represents model weights in lower precision. Post-training quantization (PTQ) applies quantization after training and can substantially reduce memory usage with varying accuracy trade-offs. For our deployment study, we convert the merged checkpoint to GGUF and use the quantization types implemented by llama.cpp\footnote{\url{https://github.com/ggml-org/llama.cpp}}, including Q8\_0 and K-quantized variants such as Q6\_K, Q5\_K\_M, Q4\_K\_M, Q3\_K\_M, and Q2\_K. These variants provide a direct way to study the trade-off between model footprint and generation quality without retraining.

\subsection{Prompt Engineering and Reasoning}

Prompt engineering is the practice of designing and optimizing natural language instructions that guide LLMs on specific tasks. Effective prompts include clear instructions, context, and examples, enabling models to perform more reliably and reducing the need for human intervention~\cite{marvin2023prompt}.

Several prompting techniques have been introduced to maximize the capabilities of LLMs. Zero-shot prompting leverages well-crafted prompts to instruct LLMs to perform tasks without explicit prior examples~\cite{kojima2022large}. Conversely, few-shot prompting integrates a small number of examples within the prompt to condition the model on the target task~\cite{brown2020language}. 


Advanced methods, such as the Chain-of-Thought and Tree-of-Thought frameworks, have emerged to explicitly guide the reasoning processes of LLMs, particularly for complex tasks. The CoT technique~\cite{wei2022chain} elicits intermediate reasoning steps and can improve performance on complex tasks, including deriving UML diagrams from user stories~\cite{li2024llm}. Variants of zero-shot CoT have also been proposed that select reasoning strategies based on model uncertainty~\cite{kumar2024enhancing}. Similarly, the ToT framework~\cite{yao2023tree} extends this concept by allowing LLMs to explore multiple reasoning paths, decomposing tasks into smaller sub-tasks. It systematically evaluates intermediate outputs to identify the most promising solutions, improving accuracy and reducing errors in complex domain modeling scenarios. Silva \emph{et al.}~\cite{silva2024application} demonstrated that applying ToT in domain modeling significantly enhanced the LLM’s ability to accurately identify and classify complex relationships in UML class diagrams, overcoming limitations encountered with simpler prompting methods.

Beyond these strategies, recent work has introduced reasoning modes that allow LLMs to perform deeper, inference-time thinking. Lee \emph{et al.}~\cite{lee2025evolving} describe approaches where models iteratively refine their reasoning through exploration and self-critique, enabling more robust problem solving.

Taken together, prompt engineering and reasoning represent complementary pathways: prompting provides lightweight mechanisms to adapt general-purpose models, while reasoning introduces deeper inference strategies that may improve structural fidelity. Both operate \emph{at inference time}, shaping model behavior on a per-request basis through the context window, whereas 
instruction tuning adapts the model weights \emph{once, ahead of deployment}, so that the adapted behavior persists across all subsequent prompts without further scaffolding. This contrast motivates our use of instruction tuning as the primary adaptation mechanism in BPMN model generation, with prompt engineering and reasoning examined as inference-time controls on top of the tuned model.

\section{Related Work}
\label{sec:rel_work}

Building upon the concepts described in Section \ref{sec:background}, we now review the related studies. We divided these studies into two sections: the first group contains rule-based and NLP/ML methods, and the second group contains recent generative AI–based approaches for automated model generation.

\subsection{Rule-based and NLP/ML-based Model Generation}

Early work focused on rule-based and NLP pipelines to convert textual process descriptions into BPMN models. Sonbol \emph{et al.}~\cite{sonbol2023machine} approached this as a machine translation task, using syntactic and semantic parsing to generate structured BPMN models, achieving up to 81\% model similarity. Similar to our evaluation approach, they employed similarity-based metrics, such as Graph Edit Distance, to compare the extracted and ground-truth models. In addition to an expert evaluation to assess their results from an expert perspective.

Similarly, Van der Aa \emph{et al.}~\cite{van2018challenges} survey how NLP techniques can support business process management tasks and outline open challenges for text-based process extraction, including coreference resolution, semantic role labelling, and handling syntactic ambiguity. Their research highlights that capturing a deeper semantic understanding is crucial for transitioning from textual artifacts to reliable process representations.

Rule-based pipelines have also shown that automatic generation of BPMN models from text is feasible in practice. Honkisz \emph{et al.}~\cite{honkisz2018concept}, for instance, propose a method that uses syntactic and semantic analysis to derive subject–verb–object constructs and gateway keywords, populate a spreadsheet-based intermediate representation, and then generate a structured BPMN model. The spreadsheet layer makes the transformation transparent and allows modelers to inspect and adjust the resulting process.

Riefer \emph{et al.}~\cite{riefer2016mining} catalogs four representative approaches, including Ghose \emph{et al.}~\cite{ghose2007process}, Sinha and Paradkar~\cite{sinha2010use}, and Epure \emph{et al.}~\cite{epure2015automatic}, alongside Friedrich \emph{et al.}~\cite{friedrich2011process} in a broader overview of pre-LLM rule-based and NLP approaches to mining process models from natural-language text. These approaches share a common architecture: POS tagging and dependency parsing, semantic resources such as WordNet and FrameNet, anaphora resolution, and handcrafted mapping rules to BPMN constructs. However, these approaches were limited by brittle assumptions about the input format, limited availability of public implementations, and a lack of standardized evaluation infrastructure. Our approach is complementary in two ways: rather than encoding the text-to-graph mapping through handcrafted linguistic rules, we learn it from paired examples; and rather than generating models for unseen text styles, we explicitly evaluate on a stratified benchmark with quantitative graph and guideline-based metrics.

\subsection{Generative AI and LLM-based Model Generation}

Recent advances in LLMs have shifted process modeling from rule-based extraction toward generation. Within this shift, approaches differ in how the LLM is adapted to the task, ranging from prompt engineering and retrieval-augmented generation to parameter-level fine-tuning and instruction tuning. 
We organize the related studies along two axes: (i) direct text-to-diagram synthesis, which maps descriptions to BPMN or a related notation; and (ii) assistant-style systems that support modeling through dialogue, retrieval, and tool integration.

On the direct-generation side, studies pair powerful LLMs with targeted components to turn text into graph-structured models. Licardo \emph{et~al.}~\cite{licardo2024method} propose combining GPT-4 with a fine-tuned BERT component to derive BPMN from structured documents and evaluate structure with relative graph edit distance. In a related vein, Nivon and Salaün~\cite{nivon2024automated} fine-tune GPT-3.5 to translate textual requirements into formal grammars and abstract syntax trees, emphasizing controllable structure, and subsequently extend the approach to BPMN optimization for execution-time refinement~\cite{nivon2024semi}. These works reflect a trend toward making intermediate structure explicit via extractors, grammars, or staged transformations. 

The closest work to ours is Soliman \emph{et al.}~\cite{soliman2025size}, who study text-to-BPMN generation by combining prompt-engineered general-purpose LLMs with fine-tuning of an open-weight model. They show that proprietary and general-purpose LLMs handle simple process descriptions reasonably well but struggle on complex cases due to schema-adherence, gateway-logic, and output-control issues, and they fine-tune Qwen2.5-Coder-14B on the MaD dataset to obtain a more reliable Description-to-DOT pipeline that subsequently converts DOT to BPMN XML through a Python-based converter. Their evaluation supports the use of DOT as a compact intermediate representation and reports efficiency gains over direct BPMN-XML generation. \GOK builds on this fine-tuning-with-DOT direction but differs in three respects relevant to deployment and evaluation. 
\begin{itemize}
  \item \textbf{Backbone size and deployability:} we instruction-tune a substantially smaller Qwen3-4B base model, targeting on-premises deployability rather than absolute size.
  
  \item \textbf{Multi-perspective evaluation:} we combine BLEU, ROUGE, and METEOR with R-GED structural fidelity, BEBoP guideline conformance, statistical testing, and a small expert review, rather than relying on a single family of metrics.
  
  \item \textbf{Deployment-time controls:} we systematically study post-training quantization and merge-time LoRA scaling as levers for trading accuracy against memory and latency without retraining.
\end{itemize}

Closely related is the recent study by Li \emph{et al.}~\cite{li2025llmbpmn}, who systematically evaluate LLM-based strategies for generating BPMN-oriented DOT representations on the MaD dataset, comparing zero-shot, zero-shot Chain-of-Thought, few-shot, few-shot Chain-of-Thought, and fine-tuning of GPT-3.5-turbo. They report that fine-tuning achieves the strongest overall performance, while few-shot Chain-of-Thought is particularly robust for structurally complex or incomplete descriptions. They evaluate their outputs using syntactic validity, node and graph-level similarity, masked-input robustness, and human judgments. \GOK is complementary in scope and method: instead of fine-tuning a closed-source API model under varied prompting regimes, we instruction-tune a compact open-weight Qwen3-4B and study its deployment-oriented variants under a single training-aligned prompt, with a multi-perspective evaluation that adds R-GED, BEBoP guideline conformance, and statistical testing alongside text similarity and expert review.

LLMs, especially the widely accessible proprietary APIs, have recently demonstrated practical instruction following and reasoning capabilities. Capitalizing on such capabilities, several studies pursue model generation via prompt-engineering approaches. Kourani \emph{et~al.}~\cite{kourani2024process} demonstrate that role prompting, knowledge injection, and multi-step verification reduce structural errors in BPMN and Petri nets by validating intermediate steps. Complementing this, assistant-style systems ground LLMs in organizational repositories so the model can reason over process artifacts and support modeling tasks; Bernardi \emph{et al.}~\cite{bernardi2024conversing} present BPLLM, a process-aware decision-support framework in which an open source LLaMA model is fine-tuned and coupled with a retrieval-augmented generation pipeline and process-aware chunking over BPMN models. Given a process repository, BPLLM answers yes/no questions about the presence of activities and sequence flows, and explains control-flow relations, demonstrating that grounding an LLM in process artifacts yields accurate, context-sensitive process explanations. 

An earlier wave of post-ChatGPT studies~\cite{grohs2023llms,klievtsova2023conversational} explored prompt-based use of general-purpose LLMs for BPM tasks such as textual process model extraction and conversational modeling assistance. These works established that LLMs can interpret process descriptions but that prompt-only outputs require expert validation, motivating the shift toward task-specific adaptation that the rest of this section reviews.


Adjacent to direct text-to-BPMN generation, several works fine-tune open-weight LLMs for related BPM tasks. Rebmann \emph{et al.}~\cite{rebmann2025potential} introduce a benchmark of five semantics-aware process mining tasks (trace-level and activity-level anomaly detection, semantic next-activity prediction, and directly-follows-graph and process-tree discovery) and show that supervised fine-tuning of open-source LLMs substantially outperforms in-context prompting on these structured tasks. Buss \emph{et al.}~\cite{buss2024processllm} present ProcessLLM, a LLaMA-2-13B model fine-tuned with LoRA in two stages:  (i) unsupervised pre-tuning on BPM literature followed by supervised tuning on BPMN-XML interpretation examples, (ii) to support analysis and optimisation of existing process models. Together with Soliman \emph{et al.}~\cite{soliman2025size}, these works converge on a consistent finding: instruction- or task-specific tuning of open-weight LLMs reliably outperforms prompt-engineered general-purpose models on structured BPM tasks. \GOK contributes to this direction with the specific focus on text-to-BPMN generation under compact, on-premises-deployable model sizes. We compare these relevant studies based on their approach, used models, adaptation technique, target language, and evaluation method in Table ~\ref{table:related_work_comparison}.
\begin{table}[!t]
\centering
\caption{Representative LLM-based BPM approaches.}
\label{table:related_work_comparison}
\scriptsize
\setlength{\tabcolsep}{4pt}
\renewcommand{\arraystretch}{1.15}
\begin{tabularx}{\textwidth}{@{}
  >{\raggedright\arraybackslash}p{2.6cm} 
  l 
  l 
  >{\raggedright\arraybackslash}X 
  >{\raggedright\arraybackslash}X 
  @{}}
\toprule
\textbf{Approach} & \textbf{Model} & \textbf{Adapt.} & \textbf{Target} & \textbf{Evaluation} \\
\midrule
Grohs \emph{et al.}~\cite{grohs2023llms} & GPT-4 & Prompt & BPMN/RPA & Qualitative \\
Klievtsova \emph{et al.}~\cite{klievtsova2023conversational} & GPT-3.5/4 & Prompt & Dialogue/extraction & Empirical \\
Licardo \emph{et al.}~\cite{licardo2024method} & GPT-4+BERT & Prompt+FT & BPMN & R-GED \\
Nivon \& Sala\"un~\cite{nivon2024automated} & GPT-3.5 & FT & Grammar/AST to BPMN & Text + structural \\
Bernardi \emph{et al.}~\cite{bernardi2024conversing} & LLaMA & FT+RAG & BPMN Q\&A & Retrieval metrics \\
Kourani \emph{et al.}~\cite{kourani2024process} & GPT-4 & Prompt & BPMN/Petri nets & Manual + structural \\
Kourani \emph{et al.}~\cite{kourani2025evaluating} & 16 LLMs & Prompt & POWL & Conformance checking \\
Rebmann \emph{et al.}~\cite{rebmann2025potential} & Open LLMs & SFT/LoRA & Process mining & Accuracy \\
Buss \emph{et al.}~\cite{buss2024processllm} & LLaMA-2 & LoRA & BPMN-XML analysis & Perplexity + expert \\
Soliman \emph{et al.}~\cite{soliman2025size} & Qwen2.5-Coder & FT & Text/DOT to XML & Accuracy + complexity \\
Li \emph{et al.}~\cite{li2025llmbpmn} & GPT-3.5 & Prompt+FT & Text to DOT & Syntactic + human \\
\midrule
\textbf{\GOK} & \textbf{Qwen3-4B} & \textbf{IT/LoRA} & \textbf{Text to DOT/XML} & \textbf{Text + R-GED + BEBoP + statistical + expert} \\
\bottomrule
\end{tabularx}
\vspace{4pt}
\raggedright
\scriptsize
\emph{Notes:} FT = fine-tuning; SFT = supervised fine-tuning; IT = instruction tuning; RAG = retrieval-augmented generation.
\end{table}

Evaluation practices coalesce around complementary lenses. For example, Ferrari \emph{et al.}~\cite{ferrari2024model} evaluates ChatGPT for UML sequence diagram generation. Their qualitative study uncovered structural inaccuracies and semantic gaps, concluding that LLM outputs require expert validation. Their study highlights the importance of qualitative analysis and human feedback when using generative AI methods, which inspired us to conduct a similar evaluation approach. Closer to our setting, Kourani \emph{et~al.}~\cite{kourani2025evaluating} present a framework that generates POWL models from text and benchmarks 16 LLMs on 20 hand-designed processes across several domains. Their evaluation is grounded in process mining, where they simulate ground-truth event logs and assess LLM outputs using conformance checking (token-based fitness and ET-Conformance precision) with a harmonic-mean quality score. 

While these generative methods show promise, many rely on proprietary APIs, handcrafted pipelines, or external data retrieval. Our approach instruction tunes an open source model with quantized adapters, offering a scalable, efficient, and privacy-preserving alternative for direct BPMN generation. The MaD dataset is used for supervised instruction tuning and held-out evaluation, but it is not accessed as an external retrieval source at inference time. The tuned model runs directly on the supplied textual description with no retrieval steps, as required for on-premise deployment.

\section{Research Method}
\label{sec:res-method}
\begin{figure}[htbp]
\centering
\includegraphics[width=0.55\linewidth]{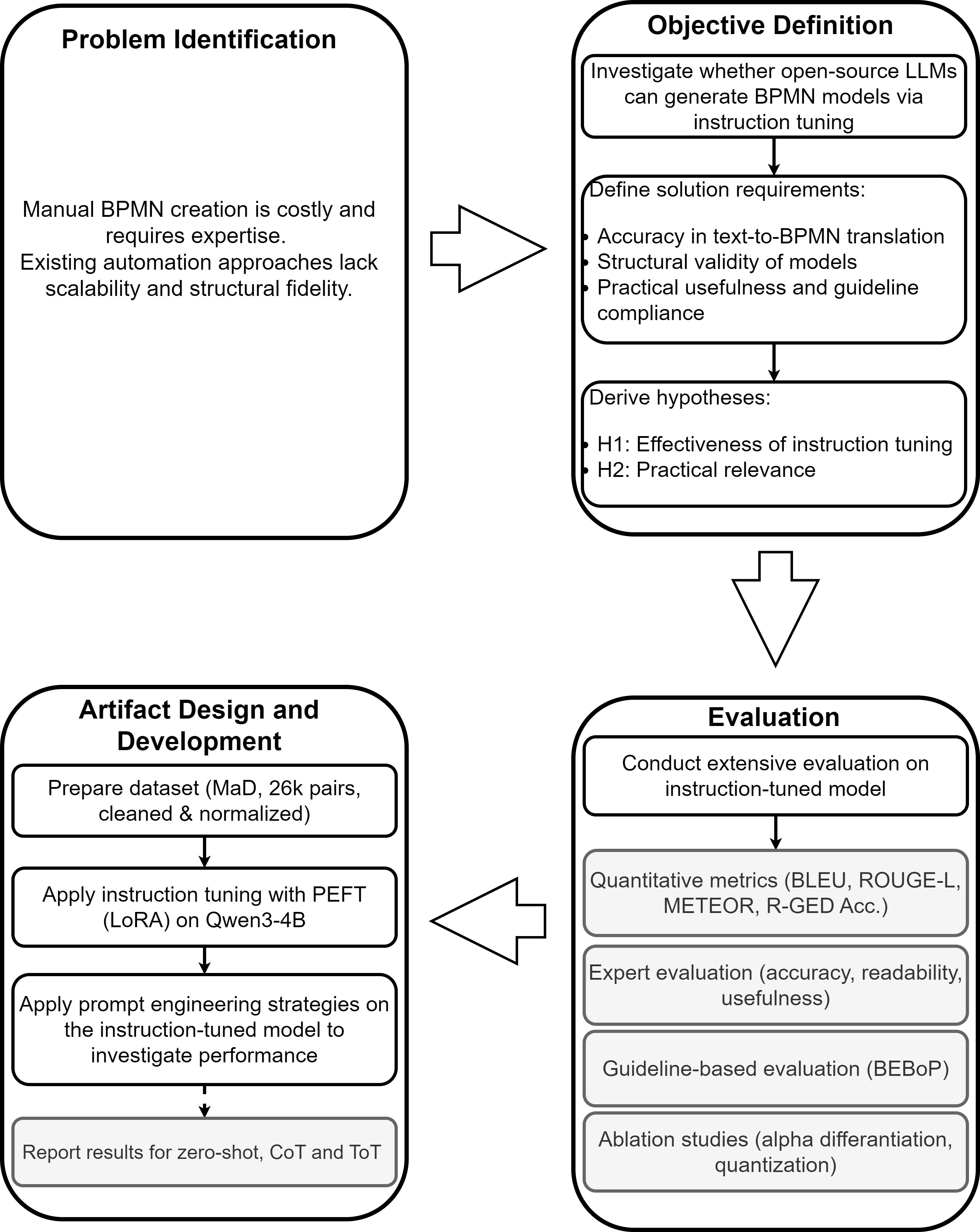}
\caption{Design Science Research methodology adopted in this study. }
\label{fig:DSR_figure}
\end{figure}

To guide the design and evaluation of our BPMN model generation method, we adopted a Design Science Research (DSR) methodology~\cite{peffers2007design}. This approach is suitable for our context as it involves creating an AI-powered artifact intended to support early-stage modeling activities. Our work has proceeded through two design–evaluation cycles in line with DSR principles. The first cycle presented in our prior work~\cite{ccelikmasat2025generating_profes} consisted of an instruction-tuned 9B-parameter open-weight model via QLoRA, and evaluated on 45 textual process descriptions using textual similarity metrics, Graph Edit Distance, and a qualitative expert study. 
This prior evaluation, however, was constrained by the relatively small number of test descriptions, the absence of guideline-based quality analysis, and a vague treatment of deployment trade-offs.
In the second cycle, we redesign the artifact as the \emph{\GOK} method based on a compact Qwen3-4B backbone and expand the evaluation with a stratified 180-instance benchmark across 15 domains, Relative Graph Edit Distance, BEBoP guideline checks, statistical tests, and an updated expert assessment. 
Across both cycles, our research method follows a consistent pattern, beginning with identifying the problem and defining objectives, then designing and developing the solution artifact, and finally evaluating its performance through both quantitative and qualitative methods. Figure~\ref{fig:DSR_figure} summarizes this process.

\subsection{Problem Identification and Objective Definition}

The manual construction of high-quality BPMN models is resource-intensive and requires specialized expertise, leading to documented quality issues even in organizations where BPMN is widely adopted~\cite{leopold2016learning}.  Textual process descriptions are widely available but are rarely translated into formal models due to this overhead. Prior efforts based on rules, pipelines, or prompt engineering have achieved partial automation but fall short in scalability, structural fidelity, and robustness (cf. Sections~\ref {sec:background} and~\ref {sec:rel_work}).

Our objective is to investigate whether open source LLMs, adapted through instruction tuning and parameter-efficient fine-tuning, can effectively automate BPMN model generation. Specifically, we examine the technical performance and the perceived practical usefulness of such models. To this end, we formulate two high-level hypotheses:

\begin{hyp}[H\ref{hyp:first}] \label{hyp:first}
Instruction tuning significantly improves the ability of LLMs to generate accurate and structurally valid BPMN models compared to untuned and proprietary baselines.
\end{hyp}

\begin{hyp}[H\ref{hyp:second}] \label{hyp:second}
The outputs of instruction-tuned LLMs are perceived as useful and adhere to established BPMN modeling guidelines, making them suitable for real-world BPM practice.
\end{hyp}

\subsection{Artifact Design and Development: \GOK}

To address these objectives, we designed and developed the \GOK method, an approach that combines instruction tuning with parameter-efficient adaptation. The artifact comprises four components: (i) dataset preparation, (ii) instruction tuning, (iii) inference, and (iv) evaluation, which are detailed in Section~\ref{sec:BPMN_generation_methodology}. The resulting models and replication package are publicly available~\cite{gokberk2025bpmn}.

\subsection{Evaluation Design}

To test H\ref{hyp:first} and H\ref{hyp:second}, we adopt a multi-perspective evaluation pipeline (detailed in Section~\ref{evaluation_pipeline}) that combines quantitative, qualitative, and guideline-based perspectives:

\begin{itemize}
    \item \textbf{Quantitative benchmarking (H\ref{hyp:first}):} BLEU, ROUGE-L, METEOR, and R-GED accuracy against open-weight and proprietary baselines, with non-parametric significance tests and bootstrap confidence intervals.

    \item \textbf{Expert evaluation (H\ref{hyp:second}):} semi-structured interviews with four BPM practitioners assessing accuracy, structural correctness, usability, and understandability.
  
    \item \textbf{Guideline-based evaluation (H\ref{hyp:second}):} automated conformance checks against a representative subset of BEBoP guidelines~\cite{corradini2018guidelines}.
  
    \item \textbf{Deployment ablations (H\ref{hyp:first} \& H\ref{hyp:second}):} controlled studies of LoRA $\alpha$-scaling and post-training quantization that bear on both effectiveness and practical relevance.
\end{itemize}

Following Baltes \emph{et al.}~\cite{baltes2025guidelines}, we report model versions, tool architecture, prompts, human validation, baselines, metrics, and limitations throughout.

\section{The Overall \GOK Method}
\label{sec:BPMN_generation_methodology}

This section details the proposed method for instruction-tuned BPMN generation, \GOK. Our approach consists of four components: 
\begin{enumerate}
    \item \textbf{Dataset preparation:} Preprocessing, cleaning, and formatting text and diagram pairs for training (Section~\ref{dataset_construction});  
    \item \textbf{Instruction tuning:} Adapting an open source LLM with parameter-efficient fine-tuning (Section~\ref{instune});  
    \item \textbf{Inference:} Generating BPMN models from unseen process descriptions using different prompting strategies (Section~\ref{inference_pipeline});  
    \item \textbf{Evaluation:} Assessing outputs through quantitative metrics, expert feedback, and guideline-based analysis (Section~\ref{evaluation_pipeline}).  
\end{enumerate}

\lstdefinestyle{llncs-listing}{
    basicstyle=\ttfamily\footnotesize,
    keywordstyle=\color{blue}\bfseries,
    commentstyle=\color{gray}\itshape,
    stringstyle=\color{orange},
    numbers=left,
    numberstyle=\tiny,
    numbersep=5pt,
    stepnumber=1,
    showstringspaces=false,
    breaklines=true,
    breakatwhitespace=true,
    captionpos=b,
    frame=single,
    framerule=0.5pt,
    xleftmargin=2em,
    xrightmargin=1em,
    aboveskip=1em,
    belowskip=1em
}

\begin{figure}[!htbp]
    \centering

    \begin{subfigure}{0.9\textwidth}
        \centering
        \fbox{%
        \begin{minipage}{0.9\textwidth}
            \ttfamily\scriptsize
            You are an expert in BPMN modeling and DOT language. Your task is to convert detailed textual descriptions of business processes into accurate BPMN model codes written in DOT language. Label all nodes with their activity names. Represent all connections between nodes without labeling the connections. Represent each node and its connections accurately, ensuring all decision points and flows are included and connected. Now, generate BPMN business process model code in DOT language for the following textual description of a business process: <BUSINESS PROCESS DESCRIPTION>
        \end{minipage}}
        \caption{Instruction template used during instruction tuning}
        \label{fig:2a}
    \end{subfigure}

    \vspace{0.3em}

    \begin{subfigure}{0.9\textwidth}
        \centering
        \begin{minipage}{0.95\textwidth}
\begin{lstlisting}[style=llncs-listing,basicstyle=\ttfamily\scriptsize]
The text below is about the account payable process. 
It starts with filing a copy for internal records. Once 
filing a copy for internal records occurs, creating a 
receiving report needs to be done. After creating a 
receiving report, one or more of the following paths should 
be executed: reporting errors to relevant team member, 
gathering the vendors invoice, doubling check a three-way 
match. Then, scheduling the payment needs to be done. 
The process is now completed. 
\end{lstlisting}
        \end{minipage}
        \caption{Sample textual input description}
        \label{fig:2b}
    \end{subfigure}

    \vspace{0.3em}

    \begin{subfigure}{0.8\textwidth}
        \centering
        \includegraphics[width=\linewidth]{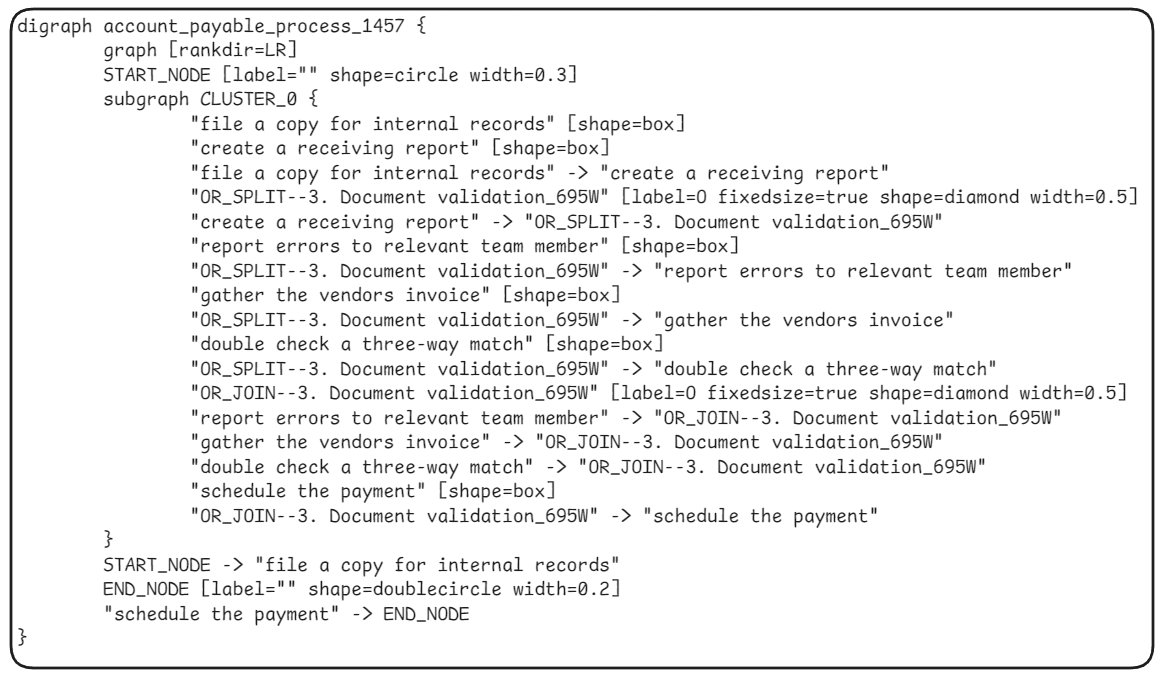}
        \caption{Model output in DOT}
        \label{fig:2c}
    \end{subfigure}

    \vspace{0.3em}

    \begin{subfigure}{0.8\textwidth}
        \centering
        \includegraphics[width=\linewidth]{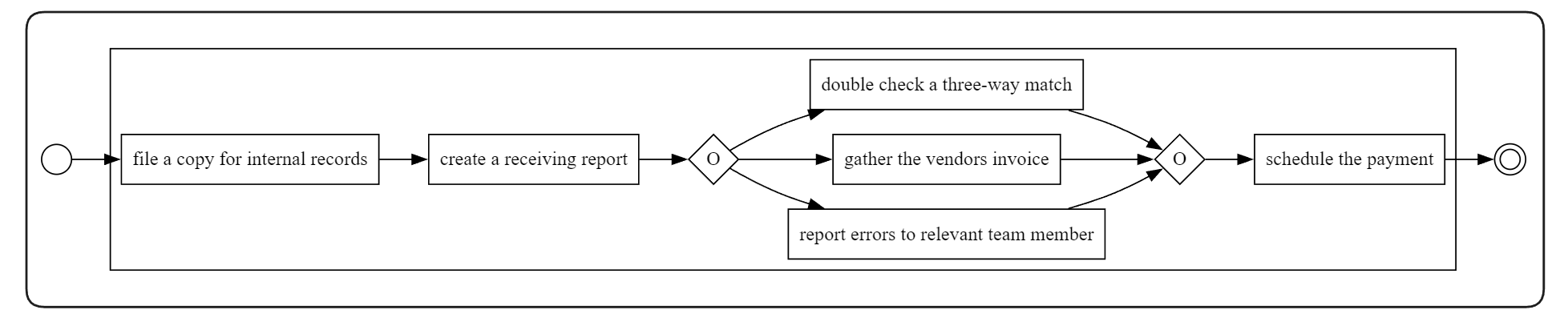}
        \caption{Rendered BPMN visualization}
        \label{fig:2d}
    \end{subfigure}

    \caption{End-to-end instruction-tuning example adapted from~\cite{ccelikmasat2025generating_profes}: (a) instruction template,
    (b) sample textual description, (c) model output in DOT, and (d) rendered BPMN model.}
    \label{fig:inst_tune}
\end{figure}

\subsection{Dataset construction}
\label{dataset_construction}

Our instruction tuning process is based on the MaD dataset~\cite{li2023mad}, which contains 30{,}000 textual business process descriptions paired with BPMN representations in DOT language\footnote{https://graphviz.org/doc/info/lang.html}, spanning 15 business domains (e.g., loan application, employee onboarding, project management).

\paragraph{Supported BPMN subset}
Our approach targets the control-flow subset of BPMN 2.0 that is captured by the MaD dataset~\cite{li2023mad}. Concretely, the supported elements are: \emph{start events} (unlabeled circle), \emph{end events} (unlabeled double circle), \emph{activities} (labeled rectangles corresponding to BPMN tasks), \emph{sequence flows} (directed edges without labels), and three \emph{gateways}: (i) parallel (AND, rendered as a \texttt{+} labeled diamond), (ii) exclusive (XOR, rendered as an \texttt{X} labeled diamond), (iii) and inclusive (OR, rendered as an \texttt{O} labeled diamond). Gateway nodes may carry descriptive names, which are preserved during DOT-to-BPMN conversion when present. This scope reflects the structure of MaD dataset and thus the following BPMN constructs are intentionally out of scope for this study and are neither present in the training data nor expected in the generated output: pools and lanes, message flows and participant choreographies, data objects and data stores, event types beyond plain start/end (intermediate, boundary, timer, message, compensation, error, signal), sub-processes and call activities, and textual annotations. 
Conditional labels on outgoing flows from XOR and OR gateways (e.g., end-state labels in the Method-and-Style convention) are not part of the structural subset we evaluate: our structural metric (R-GED) compares topology and BPMN element types rather than edge-label semantics. Where such labels do occur in the generated or reference DOT, they can still influence the string-based metrics (BLEU, ROUGE-L, METEOR), which are computed over the full DOT string.

\paragraph{Choice of DOT as the target representation}
We retain DOT as the target representation rather than converting to a richer intermediate form (e.g., BPMN 2.0 XML, POWL~\cite{kourani2025evaluating}, JSON~\cite{licardo2026bpmnassistant}, or process-tree abstract syntax trees~\cite{nivon2024automated}) for four reasons. First, MaD~\cite{li2023mad} is released in DOT, and keeping the representation aligned with the source dataset removes a transformation step that would otherwise introduce label and structural noise during training. Second, DOT is substantially more token-efficient than BPMN XML for the constructs covered in this study, which matters directly under our 2048-token context budget and fixed instruction prefix. Third, DOT has a single, unambiguous grammar with mature parsers (Graphviz, Pydot), which allows us to detect generation failures deterministically during evaluation rather than relying on downstream structural heuristics. Fourth, DOT renders directly to BPMN-style diagrams via Graphviz, so the same generated artifact is used both for quantitative evaluation and for qualitative inspection by experts.

\paragraph{Processing the data}
To ensure data quality, we applied several preprocessing steps. Using the Pydot\footnote{https://github.com/pydot/pydot} library, we removed samples with malformed DOT syntax that could not be parsed, as well as instances containing duplicate processes or disconnected components. We also excluded descriptions containing obvious typographical errors and those exceeding our maximum input token length of 2048. We adopt a 2048-token maximum input length for efficiency and coverage since MaD’s descriptions are short, and the fixed instruction adds a known, bounded prefix. When the self-attention calculations per input token are considered, increasing the cap to 4096 tokens would roughly quadruple the attention memory per layer, which would force smaller batches/throughput under our hardware budget, with limited benefit for this dataset’s length distribution, as there are not many instances with token length greater than 2048. 
These filters reduced the dataset from 30{,}000 to approximately 26{,}900 high-quality text–diagram pairs, improving consistency and structural validity. On average, the final dataset included 12.25 nodes, 13.44 edges, and 4.18 gateways per BPMN model, with textual descriptions averaging 132 words across 7.8 sentences.

For each sample (natural language description and corresponding business process model) in our filtered dataset, we dynamically created an instruction by combining the instruction template in Fig.~\ref{fig:inst_tune}\subref{fig:2a} with the sample's textual description in Fig.~\ref{fig:inst_tune}\subref{fig:2b}. The instruction template is reused from our prior work~\cite{ccelikmasat2025generating_profes}. We paired this instruction with the dataset's DOT output (Fig.~\ref{fig:inst_tune}\subref{fig:2c}) and its rendered BPMN visualization (Fig.~\ref{fig:inst_tune}\subref{fig:2d}). The dataset was split into training (80\%), validation (10\%), and test (10\%) sets and is available in our replication package~\cite{gokberk2025bpmn}. 

\subsection{Instruction Tuning Pipeline}
\label{instune}

\begin{figure}[htbp]
\centering
\includegraphics[width=0.55\linewidth]{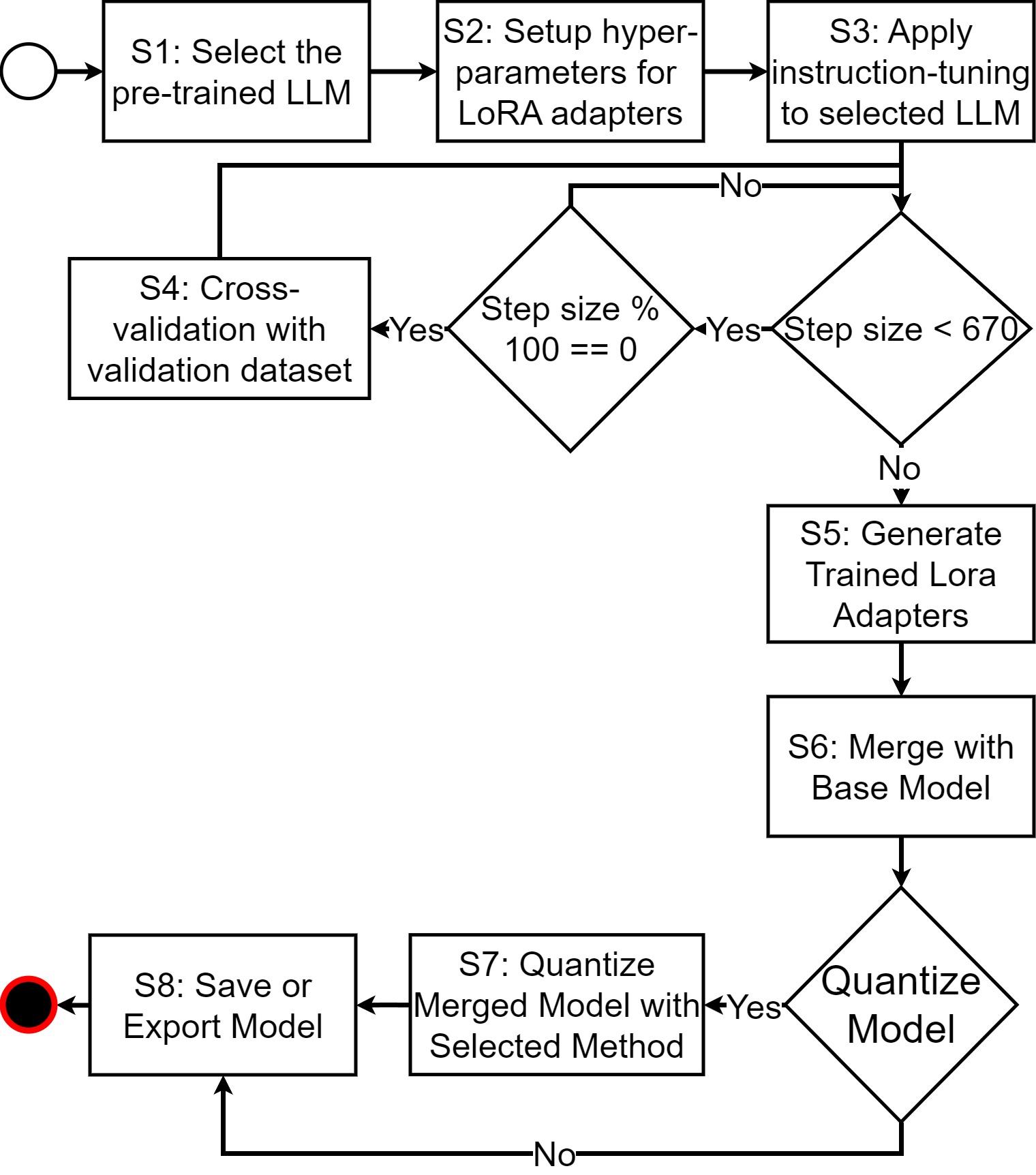}
\caption{Instruction tuning pipeline.} 
\label{fig:instune}
\end{figure}

As illustrated in Fig.~\ref{fig:instune}, the instruction tuning pipeline proceeds through eight steps. We begin by selecting Qwen3-4B \footnote{https://huggingface.co/Qwen/Qwen3-4B-Instruct-2507} for its performance-size balance, permissive licensing, and on-prem deployability (S1).
Compared to the Gemma2-9B backbone used in our prior work~\cite{ccelikmasat2025generating_profes}, Qwen3-4B offers three properties that are directly relevant for on-premises deployment: 
\begin{itemize}
  \item \textbf{Smaller footprint:} less than half ($\sim$45\%) of the active parameter count and memory footprint, which is the dominant cost at inference time.
  
  \item \textbf{Permissive licensing:} an Apache 2.0 license that removes the use-policy restrictions attached to the Gemma license family.
  
  \item \textbf{Recent base capabilities:} a July 2025 release date with corresponding improvements in base-model instruction following.
\end{itemize}
Since our prior trained model Gemma2-9B-\OLDGOK was trained on the same dataset, we keep it as a baseline in Table~\ref{table:bleu-rouge} rather than retraining it under the present pipeline, both to preserve continuity with our prior results and because a head-to-head comparison against a larger prior model is a more informative deployment signal than a same-backbone rerun.
To maintain an efficient training process while preserving output quality, we employ LoRA training and train adapters with brain floating point-16 (BF16) dtype, i.e., we do not quantize the base model during training. We enable FlashAttention 2~\cite{dao2023flashattention} and Liger Kernel~\cite{hsu2024liger} to accelerate attention and fused ops (S2).

Each dataset entry was transformed into an instruction–output pair by combining the fixed instruction template with a natural language process description and its corresponding BPMN model in DOT format. We cap the model input at 2048 tokens for coverage and efficiency. Supervised fine-tuning is then performed for one epoch over 21.5k instances using bf16, \(\texttt{lr}=2\times10^{-4}\) , \(\texttt{per\_device\_train\_batch\_size}=16\), \(\texttt{gradient\_accumulation\_steps}=2\), and LoRA \((r{=}16,\ \alpha{=}32,\ \text{dropout}{=}0.05)\). Training runs on 2\,$\times$\,L40S (48\,GB) GPUs and completes in about 150 minutes; with 21.5k instances, this corresponds to 670 optimization steps (S3).

Validation is executed every 100 steps using the validation split to monitor loss and stability (S4), and we track metrics throughout without invoking early stopping, as the schedule is a single epoch. After training, we merge the learned adapters into the base model to obtain a checkpoint (S5); crucially, we sweep a varying \(\alpha\) during merge time relative to the training \(\alpha{=}32\) before applying the actual merge. This procedure enables us to generate a set of deployment candidates with varying regularization strengths (S6). Finally, to study deployment-time accuracy/throughput/memory trade-offs, we convert the selected $\alpha{=}32$ merged checkpoint to GGUF and generate six llama.cpp quantized variants (Q8\_0, Q6\_K, Q5\_K\_M, Q4\_K\_M, Q3\_K\_M, and Q2\_K) alongside the BF16 baseline; all variants are served with vLLM v0.11.0~\cite{kwon2023efficient} under the same decoding and prompting settings used in evaluation (S7). This yields the \GOK artifact family, comprising a BF16 merged model and its quantized counterparts, all derived from a single LoRA(BF16) run on Qwen3-4B (S8). 

The resulting instruction-tuned LLMs are organised into three public HuggingFace collections and released alongside our replication package~\cite{gokberk2025bpmn}, to allow other researchers to reproduce and further test our findings:
\begin{itemize}
    \item The main tuned checkpoints reported in Table~\ref{table:bleu-rouge} are in the \emph{Text2BPMN Trained Models}  collection.\footnote{\url{https://hf.co/collections/gcelikmasat-work/text2bpmn-trained-models}}

    \item The post-training quantized variants evaluated in Table~\ref{table:quantization-impact} are in the \emph{Qwen3-4B Different Quantization GGUF} collection.\footnote{\url{https://hf.co/collections/gcelikmasat-work/qwen3-4b-different-quantization-gguf}}
  
    \item The merge-time $\alpha$ variants analysed in Table~\ref{table:alpha-sweep} are in the \emph{Qwen3-4B Different Alpha} collection.\footnote{\url{https://hf.co/collections/gcelikmasat-work/qwen3-4b-different-alpha}}
\end{itemize}


\subsection{Inference Pipeline}
\label{inference_pipeline}

\begin{figure}[htbp]
\centering
\includegraphics[width=0.70\linewidth]{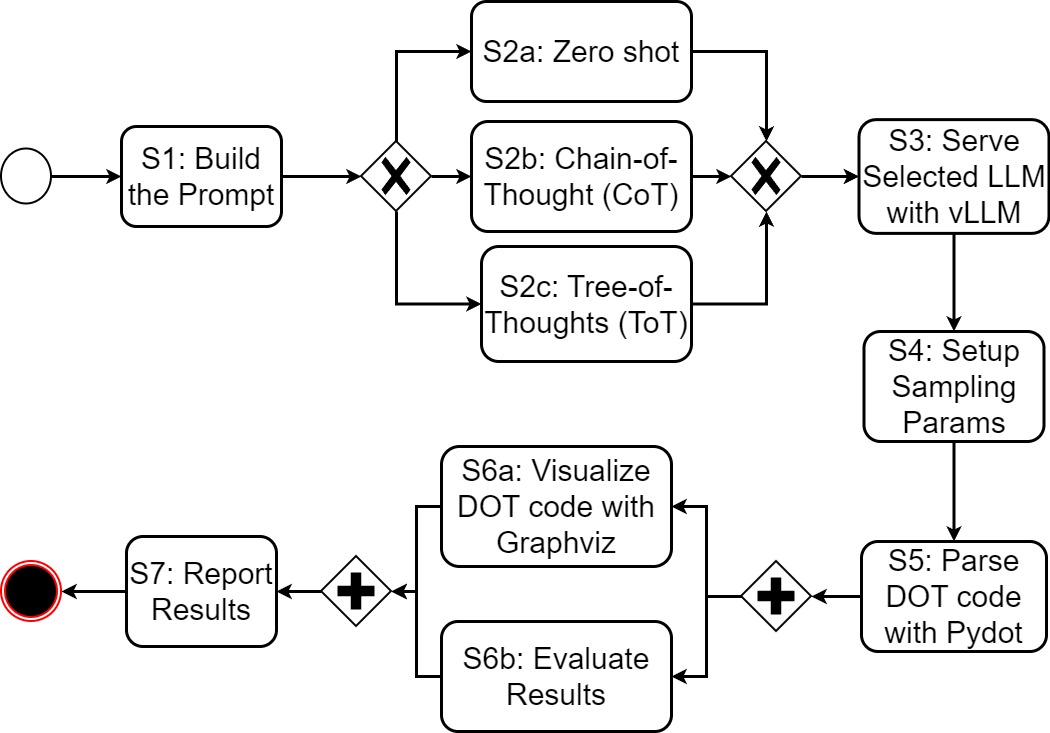}
\caption{Inference pipeline.}
\label{fig:inference}
\end{figure}

As illustrated in Fig.~\ref{fig:inference}, the inference pipeline begins with a natural language process description (S1). The description is then passed to the model under one of four prompting strategies: zero-shot, assisted zero-shot (a strengthened zero-shot prompt with a syntax example, defined below) for baselines, Chain-of-Thought, and Tree-of-Thought  (S2).

For the instruction-tuned model, we keep the same prompt format used during fine-tuning (cf. Section~\ref{dataset_construction}), which maintains alignment with its instruction–output schema and avoids unnecessary prompt complexity. For untuned baselines, we use a strengthened zero-shot prompt that includes BPMN-DOT syntax conventions and a small illustrative example, providing a common syntax reference and a fairer comparison with the tuned model. Without this scaffolding, baseline models produced incoherent or incomplete diagrams and their scores dropped substantially (quantified in Section~\ref{sec:results_and_analysis}). The full prompt is shown in Listing~\ref{fig:assisted-zero-shot}.

\begin{figure}[t]
  \centering
  \begin{minipage}{0.9\linewidth}
    \begin{lstlisting}[
      style=llncs-listing,
      numbers=none,
      basicstyle=\ttfamily\scriptsize,
      xleftmargin=1.5em,
      xrightmargin=0.5em,
      caption={Assisted zero-shot prompt given to open source models.},
      label={fig:assisted-zero-shot}
    ]
Label all nodes with their activity names. Represent all connections between nodes without labeling the connections. Represent each step and its connections accurately, ensuring all decision points and flows are included and connected. Use the following sample BPMN business process model for syntax reference:
digraph process {
graph [rankdir=LR]
START_NODE [label="" shape=circle width=0.3]
"Gather Requirements" [shape=box width=0.6]
"Design System" [shape=box width=0.6]
"Review Requirements" [shape=box width=0.6]
END_NODE [label="" shape=circle width=0.3]
START_NODE -> "Gather Requirements"
"Gather Requirements" -> "AND_SPLIT"
"AND_SPLIT" [label="+" fixedsize=true shape=diamond width=0.5]
"AND_SPLIT" -> "Design System"
"AND_SPLIT" -> "Review Requirements"
"Design System" -> "AND_JOIN"
"Review Requirements" -> "AND_JOIN"
"AND_JOIN" [label="+" fixedsize=true shape=diamond width=0.5]
"AND_JOIN" -> END_NODE
}
Now, generate BPMN business process model code in DOT language for the following textual description of a business process: <BUSINESS PROCESS DESCRIPTION>
    \end{lstlisting}
  \end{minipage}
\end{figure}

Next, the request is routed to a selected deployment variant served via vLLM  (S3). We fix the sampling parameters to temperature = 0.1, top\_p = 1.0, and a maximum generation length of 2048 tokens to encourage deterministic, diagram-only outputs within the target length budget. These settings are held constant across all models and strategies to isolate the prompting effects (S4).

The model response is captured as DOT code (S5), which we post-process to extract the diagram code if extraneous text or markup is present, then validate with Pydot. Before parsing, a minimal DOT sanitizer performs only lexical normalization—removing Markdown code fences, normalizing duplicated braces, and fixing a few recurring attribute typos (e.g., empty labels, trailing commas)—and never adds, removes, or re-routes nodes or edges. The sanitizer is applied uniformly across all models, so any structural or labeling errors in raw output are preserved and the comparison in Section~\ref{sec:results_and_analysis} is unaffected by differences in default formatting habits (e.g., Markdown code-fence wrapping).
The accepted DOT is then used in two parallel steps (S6): a) rendering to a BPMN model with Graphviz\footnote{\url{https://graphviz.org}} to be visually presentable and qualitative inspection, and b) feeding the DOT/BPMN pair into the evaluation pipeline for quantitative assessment, including textual metrics, structural fidelity. This arrangement ensures that the same artifact generated is used for both visual inspection and metric-based analysis.

\subsection{Evaluation Pipeline}
\label{evaluation_pipeline}

To test our hypotheses, we designed a multi-perspective evaluation pipeline that combines quantitative benchmarking (H\ref{hyp:first}), qualitative expert review (H\ref{hyp:second}), and guideline-based analysis (H\ref{hyp:second}). This design ensures that both the technical accuracy and the practical relevance of BPMN models generated by \GOK are systematically assessed.

\subsubsection{Quantitative Evaluation}

We evaluated our trained models against a set of open source and proprietary baselines using two types of metrics: textual similarity and structural accuracy. Together, these provide a complementary view of output quality.

\emph{Textual Similarity Metrics.} Because BPMN models are generated as DOT code, we adopted three standard metrics from text and code generation tasks:
\begin{description}
    \item[BLEU]~\cite{papineni2002bleu}: Measures precision-based n-gram overlap between generated and reference outputs.
    \item[ROUGE-L]~\cite{lin2004rouge}: Captures the longest common subsequence, reflecting both fluency and structural overlap.
    \item[METEOR]~\cite{banerjee2005meteor}: Incorporates synonym matching and word order to assess similarity at a semantic level.
\end{description}

Although not structure-aware, these metrics remain widely adopted in generative evaluation, allowing for consistent benchmarking across studies. They serve as a first-order approximation of fidelity before structural validation. 

\textbf{Relative Graph Edit Distance (R-GED):}
To capture the inherent graph structure of BPMN models, we evaluate the generated
outputs using R-GED~\cite{licardo2024method}.
sequence of node/edge insertions, deletions, and substitutions). Let
$G_{\emptyset} = (\emptyset, \emptyset)$ denote the empty graph, so that
$\mathrm{GED}(G, G_{\emptyset})$ is the cost of deleting every node and edge of
$G$; under the unit edit costs used here this equals $|V_G| + |E_G|$. Following
Licardo et al.~\cite{licardo2024method}, we define
\begin{equation} 
    \mathrm{R\text{-}GED}(G_{\text{ref}}, G_{\text{gen}}) = 1 - \frac{\mathrm{GED}(G_{\text{ref}}, G_{\text{gen}})} {\mathrm{GED}(G_{\text{ref}}, G_{\emptyset}) + \mathrm{GED}(G_{\text{gen}}, G_{\emptyset})}. 
    \label{eq:r-ged} 
\end{equation}
This yields a similarity score in $[0,1]$, where $1$ indicates a perfect
structural match. We report \emph{R-GED accuracy} as this score in percent
($0$–$100$) and compute GED values using NetworkX.\footnote{\url{https://networkx.org}} We implemented R-GED calculations into our evaluation setup in order to distinguish models that may appear syntactically similar in DOT form but differ structurally.

To disambiguate metric behavior, we explicitly state the units of comparison. BLEU, ROUGE-L, and METEOR are computed on the \emph{full DOT string} via the \texttt{nltk}~\cite{nltk2009} and \texttt{rouge\_score}~\cite{rougescore} libraries, with whitespace tokenization and no DOT-aware preprocessing. ROUGE-L additionally applies Porter stemming as in the standard \texttt{rouge\_score} implementation. Formatting differences such as attribute order, whitespace, graph name, and trailing semicolons therefore influence these metrics even when the underlying graph is structurally identical, motivating the inclusion of R-GED as a complementary structural measure. R-GED is computed on the \emph{parsed graph topology}. For this task we parse each DOT string with Pydot~\cite{pydot}, convert it to a NetworkX directed graph, and apply NetworkX's default \texttt{graph\_edit\_distance}, which treats any node as matching any other node and any edge as matching any other edge. R-GED therefore captures structural equivalence rather than label equivalence, following the computation used by Licardo \emph{et al.}~\cite{licardo2024method}. This separation of textual and structural similarity allows divergent metric behaviors (e.g.\ a high R-GED accompanied by lower text-based scores when the model recovers the correct topology with paraphrased labels) to be diagnosed cleanly.

Figure~\ref{fig:rged_example} in the appendix illustrates the size-bias-correcting behavior of R-GED on a small dataset-inspired example: the same omission of a parallel approval branch yields a different R-GED penalty depending on the size of the surrounding graph, as intended.

All models were evaluated on a fixed subset of 180 descriptions sampled from the test portion of the MaD dataset. We constructed this set by stratifying the node-count distribution within each of the 15 business domains: for every domain, we computed the 33rd and 67th percentiles of the number of BPMN nodes and then sampled 4 easy (below the 33rd percentile), 4 medium (between the 33rd and 67th percentiles), and 4 hard instances (above the 67th percentile). The subset size was chosen to balance representativeness with the practical cost of generating and evaluating outputs across multiple models. Evaluation metrics were then averaged over the sample set for comparability. This 180-instance stratified benchmark replaces the 45-instance set used in our prior work~\cite{ccelikmasat2025generating_profes}, increasing both coverage across domains and controlled variation in process complexity. All baselines and prior models in this paper, including our prior trained model, are re-evaluated on this new benchmark.

\subsubsection{Qualitative Expert Evaluation}

Quantitative metrics provide objective measures of accuracy but cannot fully capture the practical usability of BPMN models. To address H\ref{hyp:second}, we conducted a qualitative sample study~\cite{stol2018abc} with four domain experts, representing roles such as systems analysts, business analysts, workflow automation engineers, and lead business analysts, with experience ranging from 2 to 10 years (cf. Table~\ref{table:expert_demographic}). The same four experts also participated in the expert study of our prior work~\cite{ccelikmasat2025generating_profes}. However, for the present study, we conducted new semi-structured interviews in which they evaluated outputs from Qwen3-4B-\GOK, compared them against the corresponding outputs from our prior Gemma2-9B-\OLDGOK model, and assessed diagrams generated by an updated proprietary baseline (GPT-5.1, replacing the GPT-4o used previously). Due to organizational confidentiality policies, interview transcripts and written notes cannot be publicly released. Thus, we report aggregated themes and representative observations below.

\begin{table}[htbp]
    \centering
    \caption{Demographics of Experts}
    \label{table:expert_demographic}
    \begin{tabular}{llr}
       \toprule
       \multicolumn{1}{c}{ID} & \multicolumn{1}{c}{Current Role} & \multicolumn{1}{c}{Years of BPM Experience}\\
       \midrule
       E1& System Analyst &2 \\
       E2& Business Analyst & 4\\
       E3& Workflow Automation Engineer&6\\
       E4& Lead Business Analyst  & 10\\
       \bottomrule
    \end{tabular}
\end{table}

Experts were presented with generated diagrams spanning three levels of complexity: simple (fewer than 10 nodes, no parallel gateways), medium (10–20 nodes with light branching), and complex (more than 20 nodes with multiple branching patterns). They compared outputs from proprietary models (e.g., GPT-5.1), untuned open-weight baselines (Qwen2.5, Qwen2.5 Coder, Qwen3, and Qwen3 Coder), and our tuned model. The expert interviews were conducted in October 2025, following the completion of the model training phase.  Evaluations followed a rubric adapted from Ferrari \emph{et al.}~\cite{ferrari2024model}, covering four dimensions:

\begin{itemize}[label=\ding{88}]
    \item Accuracy: Faithfulness to the given textual description.
    \item Structural correctness: Logical validity of nodes, control flow, and gateways.
    \item Usability: Practical applicability with minimal post-editing effort.
    \item Understandability: Clarity and readability for business stakeholders.
\end{itemize}

Feedback was collected in semi-structured interviews and analyzed to identify recurring strengths, weaknesses, and patterns of consensus across experts. Aggregated insights are reported in Section~\ref{sec:results_and_analysis}.

\subsubsection{Guideline-Based Evaluation (BEBoP)}

To complement quantitative and qualitative assessments, we applied a rule-based evaluation using the BEBoP framework proposed by Corradini \emph{et al.}~\cite{corradini2018guidelines}. BEBoP formalizes over 50 guidelines on BPMN model clarity and provides a verifier that accepts BPMN 2.0 XML files, outputting structured JSON reports on guideline compliance.

Since our models are generated in DOT format, we implemented a converter that translates each DOT instance into BPMN 2.0 XML before submitting it to the BEBoP verifier. The converter performs a strict 1:1 structural mapping between DOT elements and BPMN constructs, summarized in Table~\ref{table:dot-bpmn-mapping}. The converter does not add, remove, merge, or re-route nodes or edges, and it does not introduce BPMN constructs absent from the DOT input. After conversion, we apply a round-trip structural check that compares the node and edge sets of the original DOT against the emitted BPMN XML; instances where this check fails are recorded as KO and excluded from BEBoP pass-rate numerators. This conservative design ensures that any BPMN diagnostic reported by BEBoP reflects the model's generated structure rather than a converter artifact.

\begin{table}[htbp]
\centering
\caption{DOT-to-BPMN 2.0 element mapping used by the converter.}
\label{table:dot-bpmn-mapping}
\footnotesize
\begin{tabular}{@{}ll@{}}
\toprule
\textbf{DOT element} & \textbf{BPMN 2.0 element} \\
\midrule
\texttt{START\_NODE} (circle)        & \texttt{bpmn2:startEvent} \\
\texttt{END\_NODE} (double circle)   & \texttt{bpmn2:endEvent} \\
Box node (labeled)                   & \texttt{bpmn2:task} \\
Diamond \texttt{[label="+"]}         & \texttt{bpmn2:parallelGateway} \\
Diamond \texttt{[label="X"]}         & \texttt{bpmn2:exclusiveGateway} \\
Diamond \texttt{[label="O"]}         & \texttt{bpmn2:inclusiveGateway} \\
Directed edge                        & \texttt{bpmn2:sequenceFlow} \\
\bottomrule
\end{tabular}
\end{table}

In line with prior studies, we focus on a selected subset of guidelines that capture key aspects of understandability, grouped by category in Table~\ref{table:bebop-selected}. For Rule 34, we follow BEBoP's automated interpretation, which checks whether diverging XOR gateways have non-empty names in the converted BPMN XML. In MaD, this name is encoded in the gateway's DOT node identifier and is copied verbatim into the BPMN \texttt{name} by our converter, which never synthesizes one. This check is therefore distinct from the labeling of outgoing branch conditions (e.g., approved''/rejected''), which our instruction format does not require the model to emit and which R-GED does not consider.

\begin{table}[htbp]
\centering
\caption{Selected BEBoP guidelines, grouped by BEBoP's own categories, with brief rationale.}
\label{table:bebop-selected}
\footnotesize
\begin{tabularx}{\textwidth}{@{}llX@{}}
\toprule
\textbf{Category} & \textbf{ID} & \textbf{Description} \\
\midrule
General    & 2  & Smaller models improve understandability and reduce cognitive load. \\
\midrule
\multirow{5}{*}{Notation}  
           & 16 & The designer should split or join sequence flows always using gateways. \\
           & 18 & Match split/join pairs to preserve structuredness. \\
           & 20 & Do not represent gateways with only one incoming and one outgoing sequence flow. \\
           & 21 & Minimize the use of inclusive (OR) gateways, which are difficult to synchronize. \\
           & 22 & Avoid dead ends at exclusive gateways; provide a default branch. \\
\midrule
\multirow{2}{*}{Labelling} 
           & 30 & Use short, meaningful verb–object labels for activities. \\
           & 34 & Use non-empty names for diverging XOR gateways. \\
\bottomrule
\end{tabularx}
\end{table}

This guideline-based analysis provides an orthogonal perspective on output quality: while text and graph metrics assess structural fidelity, BEBoP captures whether generated models conform to established modeling best practices, thereby offering additional evidence for H\ref{hyp:second} (practical relevance).
Table~\ref{table:bebop-selected} lists the selected rules with brief rationales, and Table~\ref{table:bebop-results} presents the aggregated outcomes over 180 generated models. 

\section{Results and Analysis}
\label{sec:results_and_analysis}

This section presents our experimental findings in response to the hypotheses outlined in Section~\ref{sec:res-method}. We analyze the quantitative performance of our instruction-tuned model across standard metrics and compare it with open source and proprietary LLMs (H\ref{hyp:first}). We also report qualitative insights gathered from expert evaluations to better understand the perceived usefulness and structural quality of the generated BPMN models (H\ref{hyp:second}).

\subsection{Testing H\ref{hyp:first}: Effectiveness}

\paragraph{Impact of instruction tuning on BPMN generation}

To evaluate H\ref{hyp:first}, we compared our instruction-tuned Qwen3-4B model against several baselines: our prior tuned Gemma2-9B model, our latest model's untuned counterpart, open-weight Qwen models whose parameter count is similar to or larger than ours, and a selection of state-of-the-art proprietary LLMs. 
Proprietary baselines included GPT-5.1~\cite{gpt51}, Claude-4.5 Haiku and Sonnet~\cite{claude2025}, as well as Gemini 2.5 Flash and Pro~\cite{gemini2025}. Together, these models represent the current state of the art in general-purpose LLMs and provide a strong benchmark against which to measure the effects of instruction tuning. Since all proprietary baselines support dedicated thinking modes, we evaluated them using their respective reasoning variants configured with a medium thinking-effort setting.

Each baseline was evaluated using the assisted zero-shot prompt described in Section~\ref{inference_pipeline}, while the tuned model used the same instruction format as during fine-tuning. Without such tailored prompting, untuned models performed substantially worse—approximately half the scores reported in Table~\ref{table:bleu-rouge}—highlighting that raw zero-shot performance is inadequate for generating structurally valid BPMN models in DOT notation.

\paragraph{A note on parseability}
A precondition for the metrics in Table~\ref{table:bleu-rouge} is that the generated DOT is parseable; otherwise, the corresponding R-GED contribution collapses to zero. We measured raw and post-sanitizer parse rates over the 180 generated outputs for every model in Table~\ref{table:bleu-rouge}. The pattern is consistent with the design of our sanitizer. Qwen3-4B-\GOK and its $\alpha\in\{8,16,32\}$ merge variants emit parseable DOT directly in 99.4\% of cases, with the sanitizer contributing nothing further. The mid- and high-precision GGUF variants (Q5/Q6/Q8) behave identically; aggressive low-bit settings (Q3/Q4/Q2) drop substantially in raw parseability, which is consistent with the sequence-quality degradation reported in Table~\ref{table:quantization-impact} and indicates that those settings degrade syntactically rather than only at the token level. Several proprietary baselines, by contrast, default to wrapping DOT in Markdown code fences (Claude 4.5 Haiku: 1.1\% raw, 97.8\% sanitized; Gemini 2.5 Pro: 0.0\% raw, 96.1\% sanitized), so the sanitizer improves their parseability dramatically without changing graph structure. The relative ranking in Table~\ref{table:bleu-rouge} therefore does not depend on preferential post-processing of the tuned model; if anything, it understates the gap, since the sanitizer benefits the baselines that produce wrapped output rather than the tuned model that does not.

\begin{table}[htbp]
\centering
\caption{BPMN Model Generation Textual and Structural Similarity Evaluation Results}
\label{table:bleu-rouge}
\resizebox{\textwidth}{!}{%
    \begin{tabular}{@{}lSSSS@{}}
    \toprule
    \textbf{Models} & \textbf{BLEU} & \textbf{ROUGE-L} & \textbf{METEOR} & \textbf{R-GED} \\
    \midrule
    \multicolumn{5}{l}{\textbf{Open-Weight LLMs}} \\
    Qwen2.5-7B-Instruct & 5.67 & 43.11 & 45.10 & 37.94 \\
    Qwen2.5-14B-Instruct & 6.66 & 43.01 & 42.96 & 40.00 \\
    Qwen2.5-Coder-7B-Instruct & 6.08 & 41.34 & 39.38 & 38.48 \\
    Qwen2.5-Coder-14B-Instruct & 6.76 & 43.30 & 43.99 & 42.10 \\
    Qwen3-4B-Instruct-2507 & 2.89 & 40.31 & 44.16 & 44.47 \\
    Qwen3-30B-A3B-Instruct-2507 & 6.66 & 42.28 & 44.79 & 38.68 \\
    Qwen3-Coder-30B-A3B-Instruct & 8.06 & 43.00 & 45.07 & 38.21\\
    \multicolumn{5}{l}{\textbf{Trained LLMs}} \\
    Gemma2-9B-\OLDGOK & 82.98 & 94.61 & 92.67 & 97.78 \\
    Qwen3-4B-\GOK & \cellcolor{yellow} \textbf{83.06} & \cellcolor{yellow} \textbf{94.43} & \cellcolor{yellow} \textbf{92.82} & \cellcolor{yellow} \textbf{99.44} \\
    \multicolumn{5}{l}{\textbf{Proprietary LLMs}} \\
    GPT-5.1 & 12.64 & 48.83 & 59.01 & 40.95 \\
    Gemini-2.5-Flash & 15.24 & 47.18 & 57.69 & 30.07 \\
    Gemini-2.5-Pro & 28.72 & 48.98 & 63.66 & 43.58 \\
    Claude-4.5-Haiku & 18.15 & 46.69 & 58.21 & 35.91 \\
    Claude-4.5-Sonnet & 22.56 & 49.87 & 61.37 & 41.47 \\
    \bottomrule
    \end{tabular}%
}
\end{table}

As shown in Table \ref{table:bleu-rouge}, the instruction-tuned Qwen3-4B-\GOK model attains the strongest overall performance. It surpasses every untuned open-weight and proprietary baseline on all four metrics, and against our substantially larger prior Gemma2-9B-\OLDGOK model it leads BLEU, METEOR, and R-GED accuracy while remaining statistically indistinguishable on ROUGE-L (94.43 vs.\ 94.61; overlapping 95\% confidence intervals in Table~\ref{table:macro-results-appendix}). Importantly, Qwen3-4B-\GOK attains these results with more than twice the efficiency: the earlier
Gemma2-9B-\OLDGOK model has over double the parameter count and active memory footprint. Untuned open-weight Qwen baselines stay in single-digit BLEU and around $40$–$45\%$ METEOR, and the best proprietary systems (e.g., GPT-5.1, Gemini 2.5 Pro, Claude 4.5 Sonnet) reach METEOR scores around $59$–$64\%$ and R\text{-}GED accuracies below $45\%$, substantially below both tuned models.

To confirm that these performance gaps are not due to random variation,  we first report macro means with 95\% confidence intervals on the 180-instance seed set (Table~\ref{table:macro-results-appendix}). We then ran non-parametric Friedman tests over the 14 models and the 177 diagrams, where all models produced valid outputs. For all four metrics (BLEU, ROUGE-L, METEOR, R-GED accuracy), the tests were highly significant ($p < 0.001$) with strong agreement in model rankings (Kendall's $W \ge 0.65$). The full statistics are reported in \ref{app2_statistical_tests}, (Table~\ref{table:friedman-all-metrics}).

Taken together, these very large test statistics and consistently high Kendall’s $W$ values (ranging from $0.65$ to $0.81$) demonstrate strong agreement in model rankings across all metrics. This convergence indicates that the observed performance differences are not random fluctuations but reflect a systematic and statistically reliable advantage of the instruction-tuned Qwen3-4B model over untuned open-weight and proprietary baselines.

\begin{table}[htbp]
\caption{Per-domain BPMN Model Generation Average Textual Similarity Results.}
\label{table:micro-results}

\resizebox{\textwidth}{!}{%
    \begin{tabular}{@{}lSSSS@{}}
    \toprule
    \textbf{Domain} & \textbf{BLEU} & \textbf{ROUGE-L} & \textbf{METEOR} & \textbf{R-GED} \\
    \midrule
    Account payable process                             & 84.02   & 94.69   & 93.65   & 100.0  \\
    Accounts receivable process                         & 85.62   & 95.64   & 94.17   & 100.0  \\
    Budget preparation process                          & 81.23   & 94.41   & 92.52   & 100.0  \\
    Churn-rate prevention process                       & 87.34   & 96.10   & 94.94   & 100.0  \\
    Client onboarding (marketing agency)                & 83.63   & 95.49   & 93.31   & 100.0  \\
    Content promotion process                           & 82.25   & 94.66   & 92.78   & 100.0  \\
    Customer support (ticket management)                & 62.90   & 82.53   & 79.93   & 100.0  \\
    Employee onboarding process                         & 82.87   & 94.84   & 93.14   & 100.0  \\
    Final grades submission process                     & 85.39   & 95.67   & 94.31   & 100.0  \\
    Loan application process                            & 85.49   & 95.34   & 94.01   & 100.0  \\
    Order fulfillment process                           & 85.68   & 95.48   & 94.30   & 100.0  \\
    Process for optimizing a process                    & 87.36   & 96.01   & 94.48   & 91.67  \\
    Project management process                          & 82.17   & 94.79   & 92.84   & 100.0  \\
    Purchase order workflow                             & 85.87   & 95.51   & 94.30   & 100.0  \\
    Startup due diligence (venture capitalist)          & 84.03   & 95.25   & 93.63   & 100.0  \\
    \bottomrule
    \end{tabular}%
}
\end{table}

Beyond the macro-level comparison, Table~\ref{table:micro-results} breaks down the performance of the instruction-tuned Qwen3-4B model across the fifteen domains in our test set. The model remains consistently strong on METEOR, with domain-level averages typically between $92.5\%$ and $95\%$. The only clear outlier is the \emph{Customer support (ticket management)} domain, where textual similarity drops to around 80\% METEOR and 63\% BLEU. Compared to the corpus average, a relatively high number of activities appear in the BPMN models but are only implicitly described in the instances of this domain. As a result, the tuned model often recovers the correct structure but differs more in wording and in how optional branches are grouped, which reduces the sequence overlap. 
Structurally, R-GED accuracy is $100\%$ in 14 out of 15 domains, with a single mismatch in the \emph{Process for optimizing a process} category (R-GED $=91.7\%$), suggesting that residual errors are rare and localized. 

Domain-level 95\% bootstrap confidence intervals (\ref{app2_statistical_tests}, Table~\ref{table:micro-results-appendix}) are narrow for most domains and wider only for \emph{Customer support}, indicating
that the observed variation reflects systematic differences in difficulty rather than noise.

\paragraph{ Effect of prompting strategies}

We further examined the effect of prompting strategies on our results, comparing zero-shot prompting with CoT and ToT for the instruction-tuned model. Contrary to our expectations, Table~\ref{table:prompt-eng} shows only marginal differences relative to the zero-shot instruction used in training; CoT/ToT slightly adjust sequence metrics while leaving structural accuracy essentially unchanged. This suggests the model has already internalized the knowledge needed for BPMN generation during fine-tuning, so explicit external reasoning offers little additional benefit on average.

This finding is consistent with, though contextually distinct from, the observation by Li \emph{et al.}~\cite{li2025llmbpmn} that few-shot Chain-of-Thought provides robustness gains in prompt-only and masked-input settings. In our setting, the training-aligned instruction prompt is already aligned with the model's instruction-tuned schema, so additional reasoning scaffolds add little marginal value.

\begin{table}[htbp]
\centering
\caption{Impact of Prompt Engineering Techniques on Trained Model}
\label{table:prompt-eng}
\begin{tabular}{lrrr}
\toprule
\multicolumn{1}{c}{Technique} &  \multicolumn{1}{c}{BLEU} & \multicolumn{1}{c}{ROUGE-L} & \multicolumn{1}{c}{METEOR}\\
\midrule
Zero-shot &  83.06 & 94.43 & 92.82\\
Chain-of-thought &  81.92 & 93.79 & 91.25\\
Tree-of-thoughts &  81.80 & 93.72 & 91.16\\
\bottomrule
\end{tabular}
\end{table}

\paragraph{Effect of quantization and model size}

Table~\ref{table:quantization-impact} compares the BF16 merged model with post–training–quantized GGUF variants spanning mid-precision (Q8\_0, Q6\_K, Q5\_K\_M) and more aggressive settings (Q4\_K\_M--Q2\_K). 
Mid-precision PTQ preserves BF16 accuracy on BLEU/ROUGE/METEOR while cutting memory roughly in half and delivering competitive serving performance throughput $\approx$31.04 gen tok/s, $\mathrm{TTFT_{net}}\approx$0.61\,s. Q6\_K and Q5\_K\_M offer similar accuracy and latency/throughput trade-offs with further memory savings. In contrast, heavier quantization such as Q4\_K\_M--Q2\_K degrades sequence-level quality and increases $\mathrm{TTFT_{net}}$, with limited or no throughput gains, making them less attractive for production. Runtime metrics were collected using vLLM in a closed-loop setting with a batch size of 16 parallel requests and reported as the mean of three runs. $\mathrm{TTFT_{net}}$ is computed as TTFT minus the queue time, and throughput is generated tokens per decode second.

\begin{table}[ht]
\centering
\caption{Quantization and Model Size Impact on BPMN Generation}
\label{table:quantization-impact}
\setlength{\tabcolsep}{2pt}
\resizebox{\textwidth}{!}{%
    \begin{tabular}{@{}p{2.5cm}SSSrSS@{}}
    \toprule
    \textbf{Training Mode} & 
    \textbf{BLEU} & 
    \textbf{ROUGE-L} & 
    \textbf{METEOR} & 
    \textbf{\parbox{2cm}{\centering Model \\ Size (GB)}} & 
    \textbf{\parbox{2.5cm}{\centering Throughput \\ (tokens/sec)}} & 
    \textbf{\parbox{2.5cm}{\centering Avg Time to\\1st Token (s)}} \\
    \midrule
    Full-Lora-Tuning (BF16) & 83.06 & 94.43 & 92.82 & $\sim$8.05 & 35.73 & 0.6423 \\
    Q8\_0-GGUF   & 83.08 & 94.41 & 92.83 & $\sim$4.28 & 31.04 & 0.6052 \\
    Q6\_K-GGUF   & 83.05 & 94.39 & 92.77 & $\sim$3.31 & 29.45 & 0.7249 \\
    Q5\_K\_M-GGUF   & 82.93 & 94.33 & 92.75 & $\sim$2.89 & 29.30 & 0.6710 \\
    Q4\_K\_M-GGUF   & 80.42 & 93.22 & 91.03 & $\sim$2.50 & 30.05 & 0.5502 \\
    Q3\_K\_M-GGUF   & 75.58 & 90.59 & 87.58 & $\sim$2.08 & 28.46 & 0.9269 \\
    Q2\_K-GGUF   &  7.92 & 21.72 & 22.59 & $\sim$1.67 & 28.01 & 1.7147 \\
    \bottomrule
    \end{tabular}%
}
\end{table}

\paragraph{Effect of post-hoc LoRA scaling (alpha) on model performance}

Finally, we investigated how the post-hoc merge scale of LoRA adapters affects performance. After training rank-16 adapters once, we varied the effective merge scale by changing \texttt{lora\_alpha} before merging and evaluated each merged model on the same test set. Table~\ref{table:alpha-sweep} and Fig.~\ref{fig:alpha-effect} show a sharp concave trend over $\alpha \in \{8,16,32,64\}$ with a narrow optimum and steep degradation at both extremes: too-small $\alpha$ under-applies the adaptation (slightly worse task fit), while too-large $\alpha$ can over-apply it (small generalization drop). Performance peaks in the mid-range ($\alpha=16$–$32$), where structural accuracy is highest and sequence metrics are strongest, consistent with $\alpha$ acting as a regularization/mixture knob rather than altering capacity.

\begin{table}[htbp]
\centering
\caption{Effect of merge-time LoRA alpha on model performance (rank = 16).}
\label{table:alpha-sweep}
\begin{tabular}{lrrrr}
\toprule
\textbf{Alpha ($\alpha$)} & \textbf{BLEU} & \textbf{ROUGE-L} & \textbf{METEOR} & \textbf{R-GED (\%)} \\
\midrule
8   & 41.24 & 51.13 & 58.50 & 55.07 \\
16  & 74.47 & 87.77 & 85.60 & 98.23 \\
32  & \textbf{83.06} & \textbf{94.43} & \textbf{92.82} & \textbf{99.44} \\
64  & 14.09 & 33.78 & 30.98 & 31.09 \\
\bottomrule
\end{tabular}
\end{table}

\begin{figure}[h]
\centering
\begin{tikzpicture}
\begin{axis}[
    width=0.9\linewidth,
    height=7cm,          
    xlabel={LoRA merge alpha ($\alpha$)},
    ylabel={Score},
    xtick={8,16,32,64},
    legend pos=south west,
    ymajorgrids=true,
    grid style=dashed,
    ymin=0, ymax=100
]
\addplot+[mark=o] coordinates {(8,41.24) (16,74.47) (32,83.06) (64,14.09)};
\addlegendentry{BLEU}
\addplot+[mark=triangle] coordinates {(8,51.13) (16,87.77) (32,94.43) (64,33.78)};
\addlegendentry{ROUGE-L}
\addplot+[mark=square] coordinates {(8,58.50) (16,85.60) (32,92.82) (64,30.98)};
\addlegendentry{METEOR}
\addplot+[mark=x] coordinates {(8,55.07) (16,98.23) (32,99.44) (64,31.09)};
\addlegendentry{R-GED (\%)}
\end{axis}
\end{tikzpicture}
\caption{Effect of LoRA merge alpha ($\alpha$) on evaluation metrics with rank = 16 adapters.}
\label{fig:alpha-effect}
\end{figure}

Taken together, these results provide strong evidence in support of H\ref{hyp:first}. Across all four metrics, the instruction-tuned Qwen3-4B model clearly outperforms its untuned counterpart, our prior tuned Gemma2-9B model, all
other open-weight baselines, and the strongest proprietary LLMs in our study.
The Friedman tests and bootstrap confidence intervals (\ref{app2_statistical_tests}) confirm that these gains are statistically reliable rather than random fluctuations.
Domain-level analyses show that performance remains consistently strong across the fifteen MaD process categories, with near-perfect structural accuracy in 14 out of 15 domains and only localized weaknesses in conversational process descriptions.
Ablation studies on quantization, prompt formats, and $\alpha$-based merge calibration further show that post-training choices can optimize deployment trade-offs without sacrificing accuracy. 
Overall, instruction tuning emerges as the main driver of effectiveness for open-weight LLMs in BPMN model generation,
enabling a compact model to reliably outperform significantly larger and more resource-intensive baselines.

\subsection{Testing H\ref{hyp:second}: Practical Relevance}

\paragraph{Expert perceptions of accuracy and usefulness} 

To assess H\ref{hyp:second}, we conducted structured interviews with four domain experts (cf. Table~\ref{table:expert_demographic}), who evaluated outputs from Qwen3-4B-\GOK alongside proprietary and open-weight baselines. Experts were asked to judge models across four dimensions introduced in Section~\ref{evaluation_pipeline}: accuracy, structural correctness, usability, and understandability. 

Across these dimensions, experts consistently rated the instruction-tuned Qwen3-4B model as producing clear and coherent BPMN models, outperforming the other open-source models, the proprietary GPT-5.1, and our earlier trained model Gemma2-9B-\OLDGOK. The experts described the diagrams from Qwen3-4B-\GOK as clearer, more balanced in their use of gateways and events, and closer to their own modelling style.
Labels were appropriately descriptive, control flow was logical, and most models were usable with minimal post-editing. In particular, participants praised the model’s ability to correctly capture the key steps of a process, even in moderately complex cases. However, they also identified areas for improvement:
\begin{itemize}[label=\ding{88}]
    \item The Qwen3-4B-\GOK model occasionally included redundant or overly generic node labels when textual inputs were vague.
    \item Gateway logic handling requires refinement, particularly to avoid unnecessary chaining or ambiguous loop creation.
    \item Experts expressed interest in an iterative interaction loop, where they could refine outputs by adjusting the initial prompt or engaging in a conversational exchange with the model.
    \item Lack of multilingual support was highlighted as a barrier to integration in localized workflows.
\end{itemize}

Despite these limitations, the consensus was that the Qwen3-4B-\GOK model can significantly accelerate early-stage modeling. Experts reported that the model was particularly effective for simple to medium processes. For example, E1 emphasized that the tool could rapidly produce first drafts for approval workflows, reducing manual modeling time. E2 observed that the system handled sequential tasks reliably but occasionally misused gateways. E3 emphasized that, although missing labels remained an issue, the overall structure was clear enough to refine quickly. E4 highlighted potential for onboarding and training, where automatically generated diagrams could serve as a starting point for newcomers. Experts also noted the potential for standardizing modeling practices by reducing individual variation, especially if the tool is used as an assistant in iterative workflows rather than as a fully automated solution.

Lastly, the experts’ organization expressed interest in piloting a Qwen3-4B-\GOK model-based system to automate internal documentation processes. While promising, all experts agreed that full replacement of manual modeling is premature. Instead, the tool is best suited as an augmentation mechanism to accelerate and standardize workflows.

\paragraph{Guideline adherence based on quality guidelines}

We further assessed practical relevance with the BEBoP verifier~\cite{corradini2018guidelines}, which operationalizes understandability guidelines for BPMN. Each generated DOT file was converted to BPMN~2.0 XML and checked against BEBoP, yielding structured JSON diagnostics (Well Done / Violated / Missing). Out of the 180 generated diagrams, 179 were successfully translated and verified; one case was labeled “Missing” due to a failed XML conversion. Because this missing case is rare, we omit a dedicated column and report OK/KO counts with overall pass rate in Table~\ref{table:bebop-results}. From the full set of diagnostics, we report a representative subset that admits unambiguous automated checks and jointly covers model size, control-flow structure, gateway usage, and node and gateway labeling. For each selected guideline, we compute the empirical pass rate (fraction of diagrams satisfying the rule) together with a 95\% Wilson confidence interval over the 179 verifiable diagrams. The selected rules and rationales appear in Table~\ref{table:bebop-selected}; aggregate outcomes for the instruction-tuned Qwen3-4B-\GOK model are summarized in Table~\ref{table:bebop-results}, while the corresponding confidence intervals are reported in the appendix.

\begin{table}[htbp]
\centering
\caption{Verification results on 179 generated BPMN models.}
\label{table:bebop-results}
\footnotesize
\begin{tabular}{@{}llrrr@{}}
\toprule
\textbf{ID} & \textbf{Guideline} & \textbf{OK} & \textbf{KO} & \textbf{Pass (\%)} \\
\midrule
2  & Minimize model size               & 179 & 0   & 100.00 \\
16 & Use explicit gateways             & 173 & 6   & 96.65 \\
18 & Split and join flows consistently & 175 & 4   & 97.77 \\
20 & Use meaningful gateways           & 178 & 1   & 99.44 \\
21 & Minimize inclusive OR gateways    & 81  & 98  & 45.25 \\
22 & Use default flows                 & 174 & 5   & 97.21 \\
30 & Labeling activities               & 179 & 0   & 100.00 \\
34 & Labeling XOR gateways             & 174 & 5   & 97.21 \\
\bottomrule
\end{tabular}
\end{table}

Within this selected subset, the instruction-tuned model satisfies seven of the eight rules at pass rates of 96.65\% or higher (Table~\ref{table:bebop-results}). The conformant rules span both structural soundness (explicit gateways, consistent split/join pairs, and meaningful gateway placement) and labeling (activity labels and non-empty XOR gateway names), indicating that generated diagrams adhere closely to the selected BEBoP checks. The exception is Rule~21 (minimize inclusive OR gateways) at 45.25\%, where 98 of 179 generated diagrams contain at least one OR-split. This reflects the modeling style of the MaD dataset, which frequently uses OR gateways in domains with optional or selective sub-flows. 
The tuned model has therefore inherited this pattern from the training distribution, providing a concrete example of how dataset-specific modeling conventions can be internalized during adaptation.
The remaining rule violations concentrate in three domains: customer support (ticket management), loan application, and project management. Customer support is also the per-domain textual-similarity outlier in Table~\ref{table:micro-results}, with the lowest BLEU and METEOR scores, so the guideline and text-metric lenses converge on the same comparatively harder process descriptions, where activity wording is more conversational and gateway logic more nuanced. The single R-GED outlier in Table~\ref{table:micro-results} (Process for optimizing a process) falls outside these domains, indicating that the structural and guideline weaknesses are largely localized to distinct cases rather than coinciding. Detailed Wilson 95\% confidence intervals are reported in~\ref{app2_statistical_tests} (Table~\ref{table:bebop-results-appendix}).

\section{Discussion}
\label{sec:discussion}

This section reflects on our findings, discusses their implications for both academic research and BPM practice, and outlines potential limitations and threats to validity.

\paragraph{Finding 1: Instruction tuning substantially improves BPMN generation} 
Across the 180-instance benchmark, the instruction-tuned Qwen3-4B consistently outperforms our prior tuned model, its untuned counterpart, and larger open-weight baselines, and it surpasses strong proprietary systems in structural accuracy. The tuned model attains near-perfect R-GED accuracy on our subset while leading or tying on textual metrics. Together, these results substantiate H\ref{hyp:first}, domain-specific instruction tuning is decisive for BPMN structure, reducing typical zero-shot errors in edge connectivity and control-flow logic.

\paragraph{Finding 2: Textual and structural evaluations reveal complementary insights} 
Evaluating both textual overlap (BLEU, ROUGE-L, METEOR) and structure (R-GED accuracy) uncovers divergences that a single metric family would miss. We observe textual and structural metrics diverging in both directions: high BLEU/METEOR with imperfect structure, and exact structural matches with imperfect text scores due to harmless formatting or label-ordering differences in DOT. Reporting R-GED as an accuracy-style percentage clarifies these trade-offs and motivates combined text–graph evaluation for BPMN.

\paragraph{Finding 3: Prompting and “reasoning” strategies have limited marginal value after tuning} 
For the tuned model, Chain-of-Thought and Tree-of-Thought prompts yield only marginal changes relative to the zero-shot instruction used during training; structural accuracy remains essentially unchanged. This suggests that fine-tuning already internalizes task-specific planning for BPMN, making additional explicit reasoning instructions largely unnecessary on average.

\paragraph{Finding 4: Expert involvement remains essential and aligns with guideline checks}
Practitioner feedback aligns with the quantitative picture and with our BEBoP analysis. Experts judged the tuned diagrams clear and usable with modest post-editing, while repeatedly noting generic activity labels and conservative or chained gateway patterns. These observations align with the BEBoP outcomes, highlighting high adherence to the selected control-flow and labeling checks, as well as the main gaps in gateway usage and activity wording. Together, these observations suggest that H\ref{hyp:second} holds in a pragmatic sense: instruction-tuned BPMN models are deemed useful and guideline-conforming enough to accelerate real projects. 

\paragraph{Finding 5: Post-training optimization and accuracy–efficiency trade-offs}
Post-training quantization of the merged checkpoint and merge-time scaling of the LoRA $\alpha$ values provide effective controls without additional model training. Mid-bit post-training quantization preserves near-BF16 performance on sequence metrics while substantially reducing the memory footprint; more aggressive settings clearly degrade sequence-level quality and offer limited throughput gains. Independently, the $\alpha$ sweep shows a sharp concave response: too small under-applies the adaptation, too large over-applies it; mid-range $\alpha$ yields the best balance across metrics. Together, these results may guide practitioners in tailoring fidelity versus footprint for deployment without requiring retraining.

\subsection{Implications for Research and Practice}
\label{implications_res_practice}

Our findings position instruction tuning as a practical approach to adapting open-weight LLMs for structured generation tasks, such as BPMN modeling. Rather than relying solely on retrieval pipelines or prompt engineering, instruction tuning internalizes domain constraints, resulting in a stable structure under modest computational budgets.

Evaluations of structured outputs should report both sequence overlap and graph fidelity. The divergence we observe between BLEU/ROUGE/METEOR and R-GED, underscores that surface similarity is an imperfect proxy for structural correctness. Parseability gates and failure rates should be made explicit. Guideline-based analysis complements text/graph metrics by localizing errors into actionable categories, enabling targeted ablations and future corrective strategies. Finally, deployment-time controls, such as post-training quantization and merge-time scaling, should be prioritized in methodology sections; our results show that they shape accuracy–efficiency trade-offs without requiring retraining.

For teams operating under privacy or cost constraints, compact instruction-tuned models can be deployed on-premise with near-full precision quality when using medium-bit PTQ, while mid-range merge scaling provides a simple tweak to calibrate faithfulness. The expert study and BEBoP outcomes converge on a human-in-the-loop workflow: use the model to generate first drafts quickly, then focus reviews on the most frequent issues. Organizations can also exploit these diagnostics to standardize modeling conventions and reduce variance across authors, integrating the assistant as a drafting tool rather than a fully autonomous system.

\subsection{Threats to Validity and Limitations}
\label{threats_to_validity}
Following Wohlin \emph{et al.}~\cite{wohlin2012experimentation}, we consider four categories of validity:

\emph{Construct validity.}
We formed instruction–output pairs using a fixed instruction schema and validated DOT syntax to ensure that outputs represent BPMN models rather than free text. Nonetheless, natural language descriptions can be ambiguous, so “faithfulness” may not be uniquely defined. Our structural measure (R-GED) captures graph equivalence, not semantic equivalence; two structurally identical graphs could differ in intent if the description is underspecified. Moreover, R-GED is computed on graphs reconstructed from DOT using external parsing libraries, so even small syntax defects can prevent parsing and collapse a diagram’s score to zero. We apply lightweight post-processing before parsing to reduce this brittleness, but R-GED still partly reflects the behaviour and limitations of the underlying tool chain rather than a purely model-intrinsic notion of correctness. 

\emph{Internal validity.}
Preprocessing steps may bias the dataset toward cleaner instances and can inflate structural metrics by excluding invalid generations from certain analyses. We mitigate this by evaluating all models under the same decoding settings. For guideline checks, we convert DOT to BPMN 2.0 XML using our own converter before calling BEBoP. Although we validate conversions with round-trip checks, translation errors could spuriously create or mask violations. On a separate note, the expert evaluations were conducted independently by practitioners who were not involved in model development, reducing evaluator bias.

\emph{Conclusion validity.}
While R-GED provided an effective measure of structural similarity, an R-GED of 100\% does not imply semantic equivalence in all cases. Likewise, sequence metrics are sensitive to formatting and ordering in DOT; we therefore interpret them alongside structure and guidelines. For proprietary systems, API updates and stochastic decoding can introduce run-to-run variance; we fixed decoding parameters and used a common prompt design, but residual variability remains possible. Our subset size (180 instances, stratified by domain) balances feasibility and coverage, yet smaller samples increase uncertainty around small effect sizes.

\emph{External validity.} 
Our study uses English descriptions from MaD and a fixed instruction template. Because both instruction tuning and the 180-instance benchmark are derived from MaD, the held-out evaluation assesses performance on unseen MaD instances but does not establish cross-dataset generalization. The model may therefore specialize in corpus-specific textual phrasing, BPMN conventions, gateway distributions, and DOT serialization patterns. The inherited OR-gateway behavior observed in the BEBoP analysis (Section~\ref{sec:results_and_analysis}) illustrates such dataset-specific adaptation. Consequently, dataset-level overfitting or specialization to MaD cannot be ruled out, and generalization to independent corpora, multilingual settings, other modeling notations, or uncurated enterprise documents remains unestablished.

\emph{Limitations.} 
The dataset is synthetically enhanced and may not fully reflect the diversity of real-world business documentation. Moreover, using MaD for both adaptation and evaluation means that strong benchmark performance may partly reflect recurring textual, structural, and serialization conventions specific to this corpus; transfer to an independently constructed text--process-model dataset therefore remains unknown. A stronger assessment would probe the trained model, without further adaptation, on an external corpus such as the PMo dataset introduced by Brissard \emph{et al.}~\cite{brissard2026pmr}. Such evaluation would require normalization of representation conventions because semantically equivalent process models may use different serializations, while our sequence metrics operate on the full DOT string. We also did not instruction-tune for multilingual generation, and our BEBoP analysis covers only a selected subset of guidelines.

\emph{Methodological scope.} Two methodological choices in this work are deliberate scoping decisions. First, we instruction-tune a single open-weight LLM; the choice of Qwen3-4B is motivated by on-premises deployability (Section~\ref{instune}) rather than by an ablation across candidate base models. We do, however, evaluate Qwen3-4B-\GOK against untuned open-weight LLMs with comparable active parameter counts (Table~\ref{table:bleu-rouge}), thereby isolating the effect of instruction tuning from raw model scale. What remains for future work is determining whether the same instruction-tuning recipe transfers to other base-model families, such as Gemma, Llama, or Mistral, at similar scales. Second, we commit to DOT as the intermediate representation, with the four reasons articulated in Section~\ref{dataset_construction}; we do not compare BPMN generation quality across alternatives such as BPMN 2.0 XML, POWL~\cite{kourani2025evaluating}, or grammar-based abstract syntax trees~\cite{nivon2024automated}. Our results should therefore be read as evidence of what \emph{one} compact open-weight LLM trained on \emph{one} token-efficient representation can achieve under controlled deployment constraints, and we explicitly invite replication and extension across other models and representations.

\section{Conclusions and Future Work}
\label{sec:conclusion_and_future_work}
In this study, we introduced \GOK, an instruction-tuned approach for generating BPMN process models directly from natural language descriptions. Building on Qwen3-4B, we applied parameter-efficient fine-tuning with LoRA, then produced deployable checkpoints via merge-time scaling and post-training quantization. Multi-perspective evaluation metrics, structural fidelity metric, guideline conformance, and a small expert study demonstrate that the Qwen3-4B-\GOK model achieves near-perfect structural alignment on our benchmark while remaining competitive or better on sequence metrics, outperforming untuned baselines and strong proprietary systems in terms of structure. Expert feedback and BEBoP outcomes converge: the generated diagrams are generally usable with modest post-editing, with residual issues concentrated in activity descriptions, default flows, and gateway usage. We release code, models, and evaluation scripts to support reproducibility.

Looking ahead, several lines of work follow naturally from these results: 
\begin{itemize}
  \item \textbf{Beyond BPMN:} extending instruction tuning to additional modeling languages, multi-diagram scenarios, and multilingual settings, including low-resource languages.
  
  \item \textbf{Cross-dataset generalization and data diversity:} broaden evaluation beyond MaD by probing the frozen \GOK model on independently constructed text--process-model datasets such as PMo~\cite{brissard2026pmr}, without additional instruction tuning. Such evaluation should normalize dataset-specific representation conventions and combine structure-aware with string-based metrics to distinguish semantic generalization from DOT-serialization differences.
  
  \item \textbf{Human-in-the-loop refinement:} deepening the human-in-the-loop workflow with interactive editing loops and targeted post-processing focused on the most frequent guideline violations surfaced by BEBoP.
  
  \item \textbf{Portability across base models:} applying the \GOK pipeline to additional compact open-weight LLMs of similar parameter count, to test how portable the instruction-tuning recipe is across base-model families.
  
  \item \textbf{Intermediate representations:} an empirical comparison across intermediate representations remains an open direction, since richer targets offer different trade-offs. POWL's process-tree structure facilitates conformance-checking-based evaluation, as demonstrated by Kourani \emph{et al.}~\cite{kourani2025evaluating}. Grammar-based targets such as those in Nivon and Sala\"un~\cite{nivon2024automated} impose stronger parseability guarantees. Comparing DOT against BPMN 2.0 XML, POWL, and grammar-based targets would isolate the contribution of the intermediate representation from that of the adaptation method.
\end{itemize}
In practice, our current results suggest a straightforward default for deployment: combining medium-bit PTQ to preserve near-BF16 accuracy under realistic resource constraints, enabling private, on-premise BPMN assistance that aligns closely with organizational modeling conventions.

\section*{Declarations}
\subsection*{Funding}
No funding was received for conducting this study.

\bibliographystyle{elsarticle-num}
\bibliography{bibliography}

\clearpage

\appendix

\section{R-GED on a worked example}
\label{app1_rged_example}

To illustrate how R-GED handles a structural omission of the kind frequently observed in MaD-style generation, we walk through a small dataset-inspired example. Figure~\ref{fig:rged_example} shows two diagrams: a reference $G_{\text{ref}}$ that includes a parallel approval step requiring synchronization between manager and finance branches, and a generated process $G_{\text{gen}}$ that omits the finance branch and produces a single-path flow. Nodes are abbreviated as follows: S = start, E = end, C = create requisition, M = manager approval, F = finance approval, N = notify supplier; "+" denotes a parallel gateway.

\begin{figure}[tbp]
\centering
\captionsetup{skip=8pt}
\setlength{\tabcolsep}{3pt}
\resizebox{\textwidth}{!}{%
\begin{tabular}{@{}c c@{}}
\begin{tikzpicture}[
  baseline=(current bounding box.center),
  node/.style={draw, circle, minimum size=5.5mm, inner sep=0pt},
  edge/.style={-Latex, thick},
  every node/.style={font=\small},
  scale=0.92, transform shape
]
\node[node] (s) {S};
\node[node, right=8mm of s] (c) {C};
\node[node, right=8mm of c] (as) {+};
\node[node, above right=5mm and 9mm of as] (m) {M};
\node[node, below right=5mm and 9mm of as] (f) {F};
\node[node, right=16mm of as] (aj) {+};
\node[node, right=8mm of aj] (n) {N};
\node[node, right=8mm of n] (e) {E};
\draw[edge] (s) -- (c);
\draw[edge] (c) -- (as);
\draw[edge] (as) -- (m);
\draw[edge] (as) -- (f);
\draw[edge] (m) -- (aj);
\draw[edge] (f) -- (aj);
\draw[edge] (aj) -- (n);
\draw[edge] (n) -- (e);
\end{tikzpicture}
&
\begin{tikzpicture}[
  baseline=(current bounding box.center),
  node/.style={draw, circle, minimum size=5.5mm, inner sep=0pt},
  edge/.style={-Latex, thick},
  every node/.style={font=\small}
]
\node[node] (s) {S};
\node[node, right=8mm of s] (c) {C};
\node[node, right=8mm of c] (m) {M};
\node[node, right=8mm of m] (n) {N};
\node[node, right=8mm of n] (e) {E};
\draw[edge] (s) -- (c);
\draw[edge] (c) -- (m);
\draw[edge] (m) -- (n);
\draw[edge] (n) -- (e);
\end{tikzpicture}
\\[6pt]
\small (a) Reference $G_{\text{ref}}$ (parallel approvals)
&
\small (b) Generated $G_{\text{gen}}$ (omits finance branch)

\end{tabular}%
}
\vspace{4pt}
\caption{Reference and generated BPMN diagrams illustrating an omission case.}
\label{fig:rged_example}
\end{figure}

Under unit edit costs, transforming $G_{\text{gen}}$ into $G_{\text{ref}}$ requires inserting the missing finance-approval activity, inserting the corresponding branch edges, and restoring the split/join synchronization around the parallel step. The total of these structural edits is $\mathrm{GED}(G_{\text{ref}}, G_{\text{gen}})$ in Equation~\ref{eq:r-ged}. 
The denominator, $\mathrm{GED}(G_{\text{ref}}, G_{\emptyset}) + \mathrm{GED}(G_{\text{gen}}, G_{\emptyset})$, is the cost of deleting each graph into the empty graph $G_{\emptyset}$ defined in Equation~\ref{eq:r-ged} and therefore scales with their respective sizes.
As a result, the same omission yields a larger relative penalty in this small example than it would in a process with dozens of nodes and edges: this is the size-bias correction R-GED is designed to provide. The example also makes the topological-matching property concrete — restoring the finance-approval branch is paid for as a structural cost (one inserted node, two inserted edges, restored synchronization), not as a label-mismatch cost, because R-GED's matching function does not consult node labels.

\clearpage
\section{Statistical Test Details}
\label{app2_statistical_tests}

This appendix reports the full statistical details underlying the summary results in Section~\ref{sec:results_and_analysis}.
For macro-level model comparison, we estimate 95\% confidence intervals using non-parametric bootstrap resampling over the 180-instance seed set. The confidence interval for each model is specified in Table-\ref{table:macro-results-appendix}

\begin{table}[htbp]
\centering
\caption{BPMN model generation: macro means with 95\% confidence intervals on the 180-instance seed set.}
\label{table:macro-results-appendix}

\setlength{\tabcolsep}{3pt} 

\resizebox{\textwidth}{!}{%
    \begin{tabular}{lrrrr}
    \toprule
    \textbf{Models} & \textbf{BLEU} & \textbf{ROUGE-L} & \textbf{METEOR} & \textbf{R-GED} \\
    \midrule
    \rowcolor{green!10}\multicolumn{5}{l}{\textbf{Open-weight LLMs}} \\
    Qwen2.5-7B-Instruct         &  5.67 [ 4.85,  6.82] & 43.11 [41.44, 44.89] & 45.10 [42.17, 47.51] & 37.94 [32.70, 44.14] \\
    Qwen2.5-14B-Instruct        &  6.66 [ 5.63,  8.02] & 43.01 [41.12, 44.98] & 42.96 [39.47, 46.26] & 40.00 [34.92, 45.79] \\
    Qwen2.5-Coder-7B-Instruct   &  6.08 [ 5.30,  6.95] & 41.34 [40.01, 42.61] & 39.38 [36.41, 42.09] & 38.48 [32.99, 44.32] \\
    Qwen2.5-Coder-14B-Instruct  &  6.76 [ 5.76,  8.01] & 43.30 [41.80, 44.85] & 43.99 [41.40, 46.21] & 42.10 [37.92, 46.95] \\
    Qwen3-4B-Instruct-2507      &  2.89 [ 2.16,  4.02] & 40.31 [38.70, 42.03] & 44.16 [41.52, 46.67] & 44.47 [39.38, 50.08] \\
    Qwen3-30B-A3B-Instruct-2507 &  6.66 [ 5.73,  7.72] & 42.28 [40.44, 44.16] & 44.79 [41.87, 47.27] & 38.68 [32.97, 45.25] \\
    Qwen3-Coder-30B-A3B-Instruct&  8.06 [ 6.36, 10.05] & 43.00 [41.40, 44.58] & 45.07 [40.88, 48.76] & 38.21 [32.98, 44.14] \\
    
    \rowcolor{green!10}\multicolumn{5}{l}{\textbf{Trained LLMs}} \\
    Gemma2-9B-\OLDGOK  & 82.98 [79.57, 85.09] & 94.61 [93.03, 95.49] & 92.67 [90.63, 93.85] & 97.78 [95.00, 100.00] \\
    Qwen3-4B-\GOK   & 83.06 [79.66, 85.18] & 94.43 [92.60, 95.47] & 92.82 [90.79, 94.01] & 99.44 [98.33, 100.00] \\
    
    \rowcolor{green!10}\multicolumn{5}{l}{\textbf{Proprietary LLMs}} \\
    GPT-5.1           & 12.64 [10.07, 15.55] & 48.83 [47.77, 49.87] & 59.01 [56.18, 62.04] & 40.95 [36.81, 45.79] \\
    Gemini-2.5-Flash  & 15.24 [12.63, 17.96] & 47.18 [45.97, 48.69] & 57.69 [55.94, 59.42] & 30.07 [26.05, 34.06] \\
    Gemini-2.5-Pro    & 28.72 [25.52, 31.90] & 48.98 [47.81, 50.14] & 63.66 [61.43, 65.95] &  43.58 [ 37.89,  49.63] \\
    Claude-4.5-Haiku  & 18.15 [15.06, 21.41] & 46.69 [45.58, 47.78] & 58.21 [54.87, 61.03] &  35.91 [ 31.57,  42.26] \\
    Claude-4.5-Sonnet & 22.56 [19.02, 26.29] & 49.87 [48.46, 51.40] & 61.37 [58.56, 63.94] & 41.47 [36.84, 47.10] \\
    \bottomrule
    \end{tabular}%
}
\end{table}

Per-domain micro-level results displayed on Table-\ref{table:micro-results-appendix} for our trained model are reported with bootstrapped 95\% confidence intervals across the 12 text-diagram pairs in each of the 15 domains.
For model ranking across all metrics, we apply Friedman tests with Kendall's $W$ computed over the subset of $k = 14$ models and $n = 177$ diagrams where all models produce valid outputs as seen in Table-\ref{table:friedman-all-metrics}.

\begin{table}[htbp]
\centering
\caption{Per-domain micro means with 95\% CIs for the trained model on the 180-instance seed set.}
\label{table:micro-results-appendix}

\setlength{\tabcolsep}{3pt}

\resizebox{\textwidth}{!}{%
    \begin{tabular}{lrrrr}
    \toprule
    \textbf{Domain} & \textbf{BLEU} & \textbf{ROUGE-L} & \textbf{METEOR} & \textbf{R-GED} \\
    \midrule
    Account payable process                             & 84.0 [77.9, 89.2] & 94.7 [92.6, 96.4] & 93.6 [91.4, 95.4] & 100.0 [100.0, 100.0] \\
    Accounts receivable process                         & 85.6 [80.6, 89.5] & 95.6 [93.7, 97.1] & 94.2 [91.6, 96.0] & 100.0 [100.0, 100.0] \\
    Budget preparation process                          & 81.2 [77.4, 84.2] & 94.4 [93.1, 95.4] & 92.5 [90.8, 94.0] & 100.0 [100.0, 100.0] \\
    Churn-rate prevention process                       & 87.3 [83.6, 90.2] & 96.1 [95.0, 97.0] & 94.9 [93.5, 96.1] & 100.0 [100.0, 100.0] \\
    Client onboarding (marketing agency)                & 83.6 [79.3, 87.2] & 95.5 [94.3, 96.5] & 93.3 [91.4, 94.8] & 100.0 [100.0, 100.0] \\
    Content promotion process                           & 82.2 [77.3, 86.2] & 94.7 [93.3, 95.8] & 92.8 [90.8, 94.4] & 100.0 [100.0, 100.0] \\
    Customer support (ticket management)                & 62.9 [52.1, 72.3] & 82.5 [77.2, 86.9] & 79.9 [74.4, 84.4] & 100.0 [100.0, 100.0] \\
    Employee onboarding process                         & 82.9 [77.7, 87.0] & 94.8 [93.5, 95.9] & 93.1 [91.0, 94.8] & 100.0 [100.0, 100.0] \\
    Final grades submission process                     & 85.4 [81.3, 88.6] & 95.7 [94.5, 96.6] & 94.3 [92.7, 95.6] & 100.0 [100.0, 100.0] \\
    Loan application process                            & 85.5 [80.1, 89.4] & 95.3 [94.0, 96.4] & 94.0 [92.1, 95.5] & 100.0 [100.0, 100.0] \\
    Order fulfillment process                           & 85.7 [80.7, 89.4] & 95.5 [94.1, 96.5] & 94.3 [92.7, 95.6] & 100.0 [100.0, 100.0] \\
    Process for optimizing a process                    & 87.4 [83.5, 90.2] & 96.0 [94.6, 97.2] & 94.5 [92.6, 96.0] & 91.7 [80.7, 97.0] \\
    Project management process                          & 82.2 [77.1, 86.2] & 94.8 [93.5, 95.9] & 92.8 [90.9, 94.4] & 100.0 [100.0, 100.0] \\
    Purchase order workflow                             & 85.9 [80.3, 89.8] & 95.5 [94.1, 96.5] & 94.3 [92.3, 95.8] & 100.0 [100.0, 100.0] \\
    Startup due diligence (venture capitalist)          & 84.0 [79.9, 87.9] & 95.3 [94.1, 96.4] & 93.6 [91.6, 95.2] & 100.0 [100.0, 100.0] \\
    \bottomrule
    \end{tabular}%
}
\end{table}

\begin{table}[htbp]
\centering
\caption{Friedman tests and Kendall's $W$ over $k = 14$ models and $n = 177$ diagrams.}
\label{table:friedman-all-metrics}
\begin{tabularx}{\textwidth}{Xrrrrr}
\toprule
\textbf{Metric} & $\boldsymbol{\chi^2_F}$ (df = 13) & $\boldsymbol{p}$\textbf{-value} & $\boldsymbol{W}$ & \textbf{$k$ models} & \textbf{$n$ instances} \\
\midrule
BLEU     & 1862.52 & $< 0.001$ & 0.81 & 14 & 177 \\
ROUGE-L  & 1506.76 & $< 0.001$ & 0.65 & 14 & 177 \\
METEOR   & 1727.50 & $< 0.001$ & 0.75 & 14 & 177 \\
R-GED    & 1702.32 & $< 0.001$ & 0.74 & 14 & 177 \\
\bottomrule
\end{tabularx}
\end{table}

\begin{table}[htbp]
\centering
\caption{BEBoP verification results on 179 generated BPMN models. Pass rates are reported as mean and 95\% Wilson confidence interval.}
\label{table:bebop-results-appendix}
\footnotesize
\begin{tabular}{@{}llrrr@{}}
\toprule
\textbf{ID} & \textbf{Guideline} & \textbf{OK} & \textbf{KO} & \textbf{Pass (\%) [95\% CI]} \\
\midrule
2  & Minimize model size               & 179 & 0  & 100.00 \,[97.90, 100.00] \\
16 & Use explicit gateways             & 173 & 6  & 96.65 \,[92.88, 98.45]   \\
18 & Split and join flows consistently & 175 & 4  & 97.77 \,[94.40, 99.13]   \\
20 & Use meaningful gateways           & 178 & 1  & 99.44 \,[96.90, 99.90]   \\
21 & Minimize inclusive OR gateways    & 81  & 98 & 45.25 \,[38.14, 52.57]   \\
22 & Use default flows                 & 174 & 5  & 97.21 \,[93.63, 98.80]   \\
30 & Labeling activities               & 179 & 0  & 100.00 \,[97.90, 100.00] \\
34 & Labeling XOR gateways             & 174 & 5  & 97.21 \,[93.63, 98.80]   \\
\bottomrule
\end{tabular}
\end{table}

Finally, for BEBoP guideline adherence, we report the Wilson score 95\% confidence intervals in Table-\ref{table:bebop-results-appendix} over the 179 generated BPMN models for which verification was completed successfully.

\end{document}